\documentclass{article}
\usepackage{graphicx}  
\usepackage{amsmath}   
\usepackage[compress]{cite}
\usepackage{amssymb}   
\usepackage{bm} 
\usepackage{dcolumn}
\usepackage{color}
\usepackage{mathrsfs}
\usepackage{amsfonts}
\usepackage{varioref}
\usepackage{braket}
\usepackage{subcaption}
\usepackage{comment}

\RequirePackage[colorlinks,citecolor=blue,urlcolor=magenta,linkcolor=blue]{hyperref}
\labelformat{section}{Section #1} 
\labelformat{subsection}{Section #1} 
\labelformat{subsubsection}{Section #1}
\labelformat{subsubsubsection}{Section #1}
\labelformat{equation}{Eq.~(#1)} 
\labelformat{figure}{Fig.~#1} 
\labelformat{subfigure}{Fig.~\thefigure#1} 
\labelformat{table}{Tab.~#1} 
\labelformat{appendix}{Appendix #1}
\title{A Lorentzian worldline path integral approach to Schwinger effect }
\author{Karthik Rajeev\footnote{krthkrajeev@gmail.com}
	\\
	\\
	{\small{School of Physical Sciences}}\\
	{\small{Indian Association for the Cultivation of Science, Kolkata-700032, India}}
}
\date{ }  

\begin{document}
	\maketitle
\begin{abstract}
	We demonstrate that the positive frequency modes for a complex scalar field in constant electric field (Schwinger modes), in three different gauges, can be represented as exact Lorentzian worldline path integral amplitudes. Although, the mathematical forms of the mode functions differ in each gauge, we show that a simple prescription for Lorentzian worldlines' boundary conditions dispenses the Schwinger modes in all three gauges (that we considered) in a unified manner. Following that, using our formalism, we derive the exact Bogoliubov coefficients and, hence, the particle number, \textit{without} appealing to the well known connection formulas for parabolic cylinder functions. This result is especially relevant in view of the fact that in a general electromagnetic field configuration, one does not have the luxury of closed form solutions. We argue that the real time worldline path integral approach may be a promising alternative in such non-trivial cases. We also demonstrate, using Picard-Lefschetz theory, how the so-called worldline instantons emerge naturally from relevant saddle points that are complex. 
\end{abstract}

\section{Introduction}
The presence of a strong external electromagnetic field destabilizes the vacuum of quantum field theory (QFT), inducing creation of particle pairs\cite{sauter1931verhalten,Heisenberg:1935qt,Weisskopf:1936hya,PhysRev.82.664}. The simplest and earliest known case of this pair creation process, called the Schwinger effect, corresponds to that in a constant external electromagnetic background. This phenomenon is one of the most concrete non-perturbative predictions in QFT. Although several perturbative results of QFT have been verified by impressively stringent tests, obscurities are still abundant in the non-perturbative regime. Therefore, the Schwinger effect serves as a example for theoretical explorations on the non-perturbative aspects of several fields of physics, ranging from particle physics to cosmology \cite{Ringwald:2001cp,Ringwald:2003iv,Casher:1978wy,Kharzeev:2005iz,PhysRev.183.1057,hawking1974black,Zeldovich:1971mw,Cardona:2021ovn}. 

The most satisfactory description of Schwinger effect requires a fair amount of QFT gadgetry. One way of approaching the problem is to compute the effective action of quantum electrodynamics (QED) in the presence of an external electromagnetic field (see, \cite{Dunne:2004nc} for a recent review). In general, one finds that this effective action has a non-zero imaginary part, from which pair creation rate can be computed. The real part of the effective action, on the other hand, leads to vacuum polarization. One can also use the canonical quantization approach, adapted for fields interacting with non-trivial external background\cite{popov1972pair,greiner1985quantum,haro2003pair,kluger1991pair,PhysRevD.73.065020}. In the presence of a constant electric field, for instance, one starts by computing the positive frequency modes of the appropriate field equation, in the asymptotic past. These mode functions can then be shown to evolve into linear combinations of positive and negative frequency modes in the asymptotic future. The particle number can be read off from the corresponding coefficients, called the Bogoliubov coefficients. In addition to solving the exact mode functions, several elegant methods have been employed by many investigators to study pair creation, that are especially useful when exact solutions are unavailable or overly complicated. These include, WKB approximation\cite{Martin:2007bw}, complex path methods\cite{Srinivasan:1998ty,Dumlu:2011cc}, instanton techniques \cite{PhysRevD.72.105004}, the
phase integral method\cite{PhysRevLett.104.250402}, Vlasov equation\cite{PhysRevD.45.4659,PhysRevD.82.105026,Kim:2011jw}, etc.    

The approaches to pair creation processes that use the conventional tools of QFT, although efficient, are seldom intuitive. Besides, the measurement apparatus in most experiments and the phenomenon that they observe (say, for instance, the relativistic particles in an accelerator) require localized description. On the other hand, the Fock states (and the associated S-matrix) in QFT, are inherently `extended' in nature. The worldline path integral formalism of QFT, in contrast, offers a better reconciliation with our particle based intuitions and broadens our perspective on QFT. The earliest utilization of the worldline formalism can be traced back to Feynman\cite{PhysRev.80.440,PhysRev.84.108}. In spite of being known since decades, the particular efficiency of the worldline formalism, in handling non-perturbative calculations, got its due attention only relatively recently\cite{Polyakov:1980ca,Polyakov:1987ez}. The approach was first used to find the effective action for QED in constant electromagnetic field\cite{Affleck:1981bma}, from which pair creation rate and vacuum polarizability can be calculated. This method was then later extended to more realistic scenarios, namely, the case of inhomogeneous electric field backgrounds \cite{PhysRevD.72.105004,PhysRevD.73.065028}. Numerical worldline approaches have also been helpful to study particle production in more complicated electric field configurations\cite{Gies:2005bz}. Some of the other important progresses made in the wordline path integral approach to Schwinger effect can be found in the references\cite{Ilderton:2015qda,Ilderton:2014mla,Dumlu:2010ua,Dumlu:2011rr,Torgrimsson:2017pzs,Gould:2021bre,Gould:2018ovk}      

However, the vast amount of existing literature on worldline approach to pair creation is primarily based on \textit{direct} application of Euclidean path integrals. While in some cases imaginary time is invoked in anticipation of better convergence of the path integral\cite{PhysRevD.28.2960}, in other cases it is called upon in view of, well motivated, physical insights\cite{Srinivasan:1998ty} or other strong mathematical justifications\cite{Coleman:1977py,Andreassen:2016cff}. Although the Euclidean approach seems to be immensely effectual, both in performing concrete calculations as well as in yielding useful physical insights, there are potential issues. For instance, in a general spacetime, transforming a time-like coordinate to purely imaginary values may lead to complex metric tensors. A similar issue arises for external electric field backgrounds as well. Besides, in the context of quantum cosmology, recent studies \cite{Feldbrugge:2017kzv,PhysRevD.97.023509} have shown that a certain Euclidean path integral amplitude, that was originally proposed by Hartle and Hawking to define the `wavefunction of the Universe'\cite{PhysRevD.28.2960}, has several issues when formulated in a more rigorous manner. There, an alternative approach based on real time path integrals and Picard-Lefschetz theory was proposed. We believe, that the success of the Lorentzian path integral approach in quantum cosmology, in extracting meaningful results, should be a source of inspiration to employ real time based worldline approaches to a wider class of problems in relativistic quantum theories. Into the bargain, when all is said and done, there is hardly any debate that time is very much real in the real world! This motivates us to chase the following goal: find a formalism to study pair creation in external backgrounds using both (i) the language most naturally adapted for localised particles, namely, the worldline approach to QFT, and (ii) the signature of spacetime metric that is most natural to the real world physics, namely, the Lorentzian signature. Towards this objective, in this work, we illustrate how a real time, worldline path integral formulation can be realized for the simplest case of pair creation in external background, namely, the Schwinger effect in scalar QED. It is worth mentioning, that real time worldline path integral approach, supplement by judicious numerical methods, have been used previously to study pair creation in electric field background\cite{FeldbruggeJobLeon2019}. However, to our knowledge, a concrete analytic computation of standard results in Schwinger effect using the real time formalism was hitherto unavailable. We hope that future extensions of our formalism to other pair creation processes, like Hawking radiation, may provide new insights.
  
The structure of this paper is as follows. In \ref{review_schwinger}, we give a brief review of some of the standard approaches to study Schwinger effect. It is well known that, in the canonical quantization approach, the problem of pair creation in a constant electric field can be reduced to a quantum mechanical scattering problem in an inverted harmonic oscillator (IHO) potential. Hence, as a warm up, we illustrate a real time path integral approach for finding the exact scattering wave function in an IHO potential, in \ref{review_iho}. Following that, in \ref{schwinger_effect}, we demonstrate that the positive frequency modes that describe Schwinger effect can be represented as an exact, real time, worldline path integral amplitude. We also derive the exact particle number using our formalism. Finally, in \ref{conclusion}, we summarize our findings and discuss future prospects. We have delegated the mathematical details of certain results to the Appendices. (We use the metric signature $(1,-1,-1,-1)$, henceforth)   
\section{Review of the Schwinger effect}\label{review_schwinger} 

In this section, we discuss some of the standard approaches to study Schwinger effect. For convenience, we may classify the different approaches broadly into two categories: (1) Effective action approach and (2) Canonical quantization approach. We shall now briefly review these two approaches separately.
\subsection{Effective action approach}
Shortly after the Dirac equation was discovered\cite{dirac1928quantum}, Sauter studied its solutions in the background of a constant external electric field. It was found that negative energy solutions can tunnel into positive energy solutions, which, in turn, was interpreted as creation of a hole in the Dirac's sea\cite{sauter1931verhalten}. Sauter's result clearly suggested that the Dirac equation cannot be interpreted as one describing the evolution of a one-particle wave function. Later, Heisenberg and Euler studied Dirac equation in an electromagnetic field, by treating the same in quantum field theory settings\cite{Heisenberg:1935qt}. They computed the one-loop effective action $\mathcal{S}^{(1)}_{\rm eff}[F^{\mu\nu}]$ for the electromagnetic field and found that the leading order correction to the standard action for Maxwell's Lagrangian leads to non-linear corrections to Maxwell's equation. These non-linear correction terms arise due to the interaction of electromagnetic field with the vacuum fluctuations. In their work, Heisenberg and Euler also pointed out that $\mathcal{S}^{(1)}_{\rm eff}[F^{\mu\nu}]$ has an imaginary part, although they did not compute it explicitly. Subsequently, in his seminal paper, Schwinger calculated the imaginary part of $\mathcal{S}^{(1)}_{\rm eff}[F^{\mu\nu}]$ for a constant electromagnetic field exactly\cite{PhysRev.82.664}. A non-zero imaginary part for the effective action indicates decay of the vacuum against pair creation process. In the specific case of constant external electric field of magnitude $E$, the Schwinger's formula for the vacuum decay rate $\Gamma(F^{\mu\nu})$ reads:
\begin{align}\label{Gamma_E}
	\Gamma(F^{\mu\nu})\equiv 2\hbar^{-1}\mathcal{L}^{(1)}_{\rm eff}=\frac{(2s+1)}{(2\pi)^3}\left(\frac{qE}{\sqrt{c}\hbar}\right)^2\sum_{n=1}^{\infty}\frac{e^{\frac{2\pi i (n-1)}{(2s+1)}}}{n^2}\exp\left[-\left(\frac{\pi m^2c}{|qE|\hbar}\right)n\right]
\end{align} 
where, $q$ and $m$ are, respectively, the mass and charge, and the spin $s=0$ corresponds to scalar QED and $s=1/2$ corresponds to spinor QED. In this work, we shall be focusing only on the scalar QED case, henceforth. 

The vacuum decay rate for scalar QED in an arbitrary electromagnetic field can also be written as a Euclidean quantum mechanical path integral\cite{PhysRev.84.108,PhysRev.80.440}:
\begin{align}\label{Gamma_E_Euc_int}
	\Gamma(F^{\mu\nu})={\rm Im}\left[\int_{0}^{\infty}\frac{dT}{T}e^{-\frac{mc^2}{2}T}\int_{x^{\mu}(0)=x^{\mu}(T)}\mathcal{D}[x^{\mu}]\exp\left\{-\frac{1}{\hbar}\int_{0}^{T}\left(\frac{m}{2}\delta_{\mu\nu}\dot{x}^{\mu}\dot{x}^{\nu}-i\frac{q}{c}A_{\mu}\dot{x}^{\mu}\right)d\tau_{\rm E}\right\}\right]
\end{align}
where, $\tau_{E}$ is the Euclidean proper time, $A_{\mu}$ is the electromagnetic gauge field and the path integral is over periodic trajectories satisfying $x^{\mu}(0)=x^{\mu}(T)$. Since the above path integral involves sum over worldlines of a relativistic particle, the formalism based on such path integrals is called worldine path integral formalism, or Euclidean worldine path integral formalism, when the proper time is taken to be purely imaginary as in \ref{Gamma_E_Euc_int} (see \cite{Edwards:2019eby} for a recent review).  Many years after Schwinger's work, Affleck et al. reproduced \ref{Gamma_E} for a constant electric field using the Euclidean worldline path integral approach\cite{Affleck:1981bma}. When exact computations are not possible, one resorts to the saddle point approximation of \ref{Gamma_E_Euc_int} to calculate the vacuum decay rate. The dominant contributions to $\Gamma(F^{\mu\nu})$ come from classical trajectories that are periodic in the Euclidean proper time $\tau_{E}$. Such classical trajectories have been dubbed the `worldline instantons'\cite{PhysRevD.72.105004,PhysRevD.73.065028}. 

\subsection{Canonical quantization approach}
A major drawback of the effective action approach, in the study of pair creation processes, is that one cannot get the explicit expression for the number or rate of pairs produced. To this end, one usually resorts to the canonical quantization approach. Consider the Klein-Gordon equation for a complex scalar field interacting with an electromagnetic field background described by gauge field $A_{\mu}$:
\begin{align}\label{KG_with_em}
	\left[\left(i\hbar\partial_{\mu}+\frac{q}{c}A_{\mu}\right)^2-m^2\right]\phi=0
\end{align} 
To solve the above equation, we have to make a gauge choice. For the case of a constant electric field $E$ along, say, the $x$-direction, there are different choices found in literature, out of which we shall be focusing on three most popular ones, namely, the time dependent gauge: $A^{(1)}_{\mu}=(0,E ct,0,0)$, the space dependent gauge: $A^{(2)}_{\mu}=(-E x,0,0,0)$ and the lightcone gauge: $A^{(3)}_{\mu}=\frac{E}{2}(-ct-x,ct+x,0,0)$. The standard practice is to solve for a complete set of positive frequency modes of \ref{KG_with_em}, which are orthonormal under the Klein-Gordon inner product. The presence of electric field brings in certain non-trivial aspects which, in turn, leads to pair creation. However, this manifests in seemingly different manners in each choice of the gauge. Hence, we shall briefly look at the problem in the three gauges separately. 

\subsubsection{The time dependent gauge}\label{review_tdependent}
In the time dependent gauge, we can seek solutions of the form:
\begin{align}
	\phi(x^{\mu})\propto e^{\frac{i}{\hbar}\bm{k}.\bm{x}}\xi^{(1)}_{\bm{k}}(t)
\end{align}
where, $\xi^{(1)}_{\bm{k}}(t)$ satisfies the following differential equation:
\begin{align}\label{KG_tdependent}
	-\frac{\hbar^2}{2m}\frac{d^2\xi^{(1)}_{\bm{k}}}{d(ct)^2}-\frac{1}{2}m\omega^2\left(ct-\frac{k_xc}{qE}\right)^2\xi^{(1)}_{\bm{k}}=\epsilon_{k_{\perp}}\xi^{(1)}_{\bm{k}},
\end{align} 
where we have defined
\begin{align}
\omega\equiv\frac{|qE|}{mc},\quad k_{\perp}^2\equiv k_y^2+k_z^2\quad\textrm{and}\quad\epsilon_{k_{\perp}}\equiv\frac{(k_{\perp}^2+m^2c^2)}{2m}
\end{align}
The exact solutions to \ref{KG_with_em}, that correspond to the positive (and negative) frequency modes in the asymptotic past $U^{(1)}_{\bm{k}}(x^{\mu})$ (and $U^{(1)*}_{\bm{-k}}(x^{\mu})$), which we shall call past positive(negative) modes, for short, can be written in terms of parabolic cylinder functions\cite{padmanabhan1991quantum}:
\begin{align}\label{U_TD_gauge}
	U^{(1)}_{\bm{k}}(x^{\mu})\propto e^{\frac{i}{\hbar}\bm{k}.\bm{x}}D_{\nu_{\bm{k}}}\left[e^{\frac{3\pi i}{4}}\sqrt{\frac{2m\omega}{\hbar}}\left(ct-\frac{k_xc}{qE}\right)\right]\quad;\quad\nu_{\bm{k}}\equiv \left(-\frac{1}{2}+\frac{i\epsilon_{k_{\perp}}}{\hbar\omega}\right).
\end{align}
Similarly, one can also find the positive (and negative) frequency solutions say $V^{(1)}_{\bm{k}}(x^{\mu})$ (and $V^{(1)*}_{-\bm{k}}(x^{\mu})$) in the asymptotic future. Then, using the properties of parabolic cylinder functions, one can show that $U^{(1)}_{\bm{k}}(x^{\mu})$ can be written as a linear combination of $V^{(1)}_{\bm{k}}(x^{\mu})$ and $V^{(1)*}_{-\bm{k}}(x^{\mu})$, such that:
\begin{align}
	U^{(1)}_{\bm{k}}(x^{\mu})=\alpha_{\bm{k}} V^{(1)}_{\bm{k}}(x^{\mu})+\beta_{\bm{k}} V^{(1)*}_{-\bm{k}}(x^{\mu})
\end{align}
Hence, the vacua, and consequently the associated particle definition, corresponding to the modes $U^{(1)}_{\bm{k}}(x^{\mu})$ and $V^{(1)}_{\bm{k}}(x^{\mu})$ are inequivalent. Using the principles of canonical quantization, we can show that the vacua corresponding to $U^{(1)}_{\bm{k}}(x^{\mu})$ is populated by a distribution $n_{\bm{k}}$ of $V^{(1)}_{\bm{k}}-$type particle pairs, where
\begin{align}\label{particle_t_gauge}
	n_{\bm{k}}=|\beta_{\bm{k}}|^2=\exp\left[-\frac{c(k^2_{\perp}+m^2c^2)}{|qE|\hbar}\right].
\end{align}    

Note that \ref{KG_tdependent} can be interpreted as a fixed-energy Schr\"{o}dinger equation, for a particle in an inverted harmonic oscillator (IHO) potential, with the energy being positive. Therefore, the problem of finding the positive/negative frequency modes reduces to a quantum mechanical scattering problem in IHO potential. Specifically, the solution of \ref{KG_tdependent} corresponding to $U^{(1)}_{\bm{k}}(x^{\mu})$ can interpreted as the `wave function' for a fictitious quantum mechanical particle incident towards the `left' (i.e., with respect to the time axis) from $t\rightarrow\infty$. The particle number $n_{\bm{k}}$, in this picture, gets the interpretation of the ratio of reflection probability $\mathcal{P}_{r}$ to transmission probability $\mathcal{P}_{t}$. This ratio, in turn, has been reproduced by several elegant methods, without resorting to exact solutions. These include complex path methods\cite{Srinivasan:1998ty}, instanton techniques\cite{Kim:2000un}, etc. 

\subsubsection{Space dependent gauge}\label{review_xdependent}
In the space dependent gauge, we can look for solutions of the form:
\begin{align}
	\phi(x^{\mu})\propto e^{-\frac{i}{\hbar}k_t t}e^{\frac{i}{\hbar}(k_yy+k_zz)}\xi^{(2)}_{k_t,\bm{k}_{\perp}}(x)
\end{align}
where $k_t>0$ and $\xi^{(2)}_{k_t,\bm{k}_{\perp}}(x)$ satisfies the following differential equation:
\begin{align}\label{KG_xdependent}
	-\frac{\hbar^2}{2m}\frac{d^2\xi^{(2)}_{k_t,\bm{k}_{\perp}}}{dx^2}-\frac{1}{2}m\omega^2\left(x-\frac{k_t}{qE}\right)^2\xi^{(2)}_{k_t,\bm{k}_{\perp}}=-\epsilon_{k_{\perp}}\xi^{(2)}_{k_t,\bm{k}_{\perp}}.
\end{align}
The exact solutions to the above equation can be written in terms of parabolic cylinder functions. For instance, the positive frequency solution $U^{(2)}_{\bm{k}}(x^{\mu})$ that corresponds to purely left moving wave along the $x-$axis as $x\rightarrow\infty$ is given by\cite{padmanabhan1991quantum}:
\begin{align}
U^{(2)}_{k_t,\bm{k}_{\perp}}(x^{\mu})\propto e^{-\frac{i}{\hbar}k_t t}e^{\frac{i}{\hbar}(k_yy+k_zz)}D_{\nu^*_{\bm{k}}}\left[e^{\frac{\pi i }{4}}\sqrt{\frac{2m\omega}{\hbar}}\left(x-\frac{k_tc}{qE}\right)\right].
\end{align}
The above mode function is then interpreted as describing an antiparticle particle incident towards the right from $x\rightarrow-\infty$, which, in turn, gets reflected and transmitted in the electromagnetic potential. One finds that in addition to a non-zero transmission coefficient $\mathcal{T}$, the reflection coefficient $\mathcal{R}$ is greater than unity, which is attributed to particle production. An exact calculation yields:
\begin{align}
	\mathcal{T}=\exp\left[-\frac{c(k^2_{\perp}+m^2c^2)}{|qE|\hbar}\right]&&\mathcal{R}=1+\exp\left[-\frac{c(k^2_{\perp}+m^2c^2)}{|qE|\hbar}\right]
\end{align} 

\subsubsection{The lightcone gauge}
This is the choice of gauge in which the mode functions have the simplest form\cite{Srinivasan:1998ty,Srinivasan:1999ux}. Let us first define the lightcone coordinates $u$ and $v$ as:
\begin{align}
	u=ct-x\quad;\quad v=ct+x.
\end{align}
Then, we can look for solutions to \ref{KG_with_em} of the following form:
\begin{align}
	U^{(3)}_{k_{v},\bm{k}_{\perp}}(x^{\mu})=e^{-\frac{i}{\hbar}\frac{k_v}{2} u}e^{\frac{i}{\hbar}(k_yy+k_zz)}\xi^{(3)}_{\bm{k}}(v)
\end{align}
where, $\xi^{(3)}_{\bm{k}}(v)$ satisfies the following linear differential equation:
\begin{align}
	\frac{d\xi_{\bm{k}}^{(3)}}{dv} \left(v-\frac{c k_v}{qE}\right)+\xi_{\bm{k}}^{(3)} \left(\frac{1}{2}-\frac{i \epsilon_{k_\perp} }{\hbar \omega}\right)=0
\end{align}
The solutions $U^{(3)}_{k_{v},\bm{k}_{\perp}}(x^{\mu})$, which are positive frequency modes with respect to $v$ in the limit $v\rightarrow-\infty$ is given by:
\begin{align}\label{U3_modes_review}
	U^{(3)}_{k_{v},\bm{k}_{\perp}}(x^{\mu})\propto e^{-\frac{i}{2\hbar}k_v u}e^{\frac{i}{\hbar}(k_yy+k_zz)}\left(1-\frac{qE}{ck_v}v\right)^{\nu_{\bm{k}}}
\end{align}  
Note that the above solution diverges at $v=ck_v/qE$, which turns out to be an asymptote for the projection of appropriate classical solution in the $t-x$ plane. In this gauge, the number $n_{\bm{k}}$ is then read off from the probability for quantum mechanical tunnelling to the classically forbidden region $v>ck_v/qE$. Specifically,
\begin{align}
	n_{\bm{k}}=\left|\frac{U^{(3)}_{k_{v},\bm{k}_{\perp}}\left(v>\frac{ck_v}{qE}\right)}{U^{(3)}_{k_{v},\bm{k}_{\perp}}\left(v<\frac{ck_v}{qE}\right)}\right|^2=\exp\left[-\frac{c(k^2_{\perp}+m^2c^2)}{|qE|\hbar}\right]
\end{align}
which is consistent with the results found using the other gauges. It is worth mentioning that mode functions in the lightcone gauge have a striking similarity to the scalar field modes near a black hole horizon. However, there are also some important distinctions, a discussion of which can be found, for instance, in reference\cite{Srinivasan:1998ty}.

One of the main goals of this paper is to show that, using a unifying principle, all the three positive frequency mode functions $U^{(1)}_{\bm{k}}$, $U^{(2)}_{k_t,\bm{k}_{\perp}}$ and $U^{(3)}_{k_v,\bm{k}_{\perp}}$ can be represented as real time worldline path integral amplitudes. Before doing that, in light of the relevance of IHO system in our discussion so far and as a warm up exercise, in the next section, we shall review a rather unconventional approach for studying quantum mechanical scattering problem in an IHO potential. 

\section{Quantum scattering in the IHO potential}\label{review_iho}
Scattering in a static potential is usually studied using the time-independent Schr\"{o}dinger equation. In the case of IHO, the appropriate equation is given by:
\begin{align}
	-\frac{\hbar^2}{2m}\frac{d^2\psi}{dx^2}-\frac{1}{2}m\omega^2x^2\psi=\epsilon\psi
\end{align}
One then proceeds to solve the above equation with the boundary condition appropriate for a scattering problem. For instance, the wave function $\psi^{(L)}_{\epsilon}$ that corresponds to a particle incident towards the left from $x\rightarrow\infty$, which is then reflected and transmitted in the IHO potential, is found to be \cite{padmanabhan1991quantum}:
\begin{align}\label{psi_L_first_appear}
	\psi^{(L)}_{\epsilon}(x_1)\propto D_{\nu}\left(e^{\frac{3\pi i}{4}}\sqrt{\frac{2m\omega}{\hbar}}x_1\right)\quad;\quad \forall x_1\in\mathbb{R}
\end{align}
where,
\begin{align}
	\nu=\frac{i\epsilon}{\hbar\omega}-\frac{1}{2}
\end{align}
The path integral formulation of quantum scattering problem is often not given its due share in most textbook discussions. In the case of IHO, for instance, one would expect that $\psi^{(L)}_{\epsilon}(x_1)$ can be expressed as a path integral, over paths that satisfy a certain boundary condition. We shall shortly see, that this can be done explicitly. Prior to that, let us briefly explore the classical dynamics of IHO system, keeping in mind that we are ultimately interested in the scattering problem. 

\subsection{Classical dynamics in IHO}
The classical equation of motion for IHO is given by:
\begin{align}\label{IHO_EOM}
	\ddot{x}-\omega^2x=0
\end{align}
The general solution to the above equation can be written as:
\begin{align}
	x(t)=x_{-} e^{-\omega t}+x_{+}e^{\omega t} 
\end{align}
where, $x_{\pm}$ are arbitrary constants. Be reminded that we are in pursuit of a path integral representation for the wave function $\psi^{(L)}_{\epsilon}(x_1)$. What are the boundary conditions appropriate to this problem? Since $\psi^{(L)}_{\epsilon}(x_1)$ is the quantum mechanical amplitude at the point $x_1$, it is clear that the final boundary condition should be:
\begin{align}\label{final_condition}
	x(T)=x_1
\end{align}
where the instant $T$ corresponds to the final time. We have to impose one more boundary condition, namely, that at the initial time, say, $t=T_0$. The requisite boundary condition at $T_0$ should be such, that imposing the same \textit{unambiguously} dispenses the classical solution corresponding to a particle incident towards the left from $x\rightarrow\infty$.

\begin{figure}[t]
	\centering
	\begin{subfigure}[b]{0.4\textwidth}
		\includegraphics[width=1\textwidth]{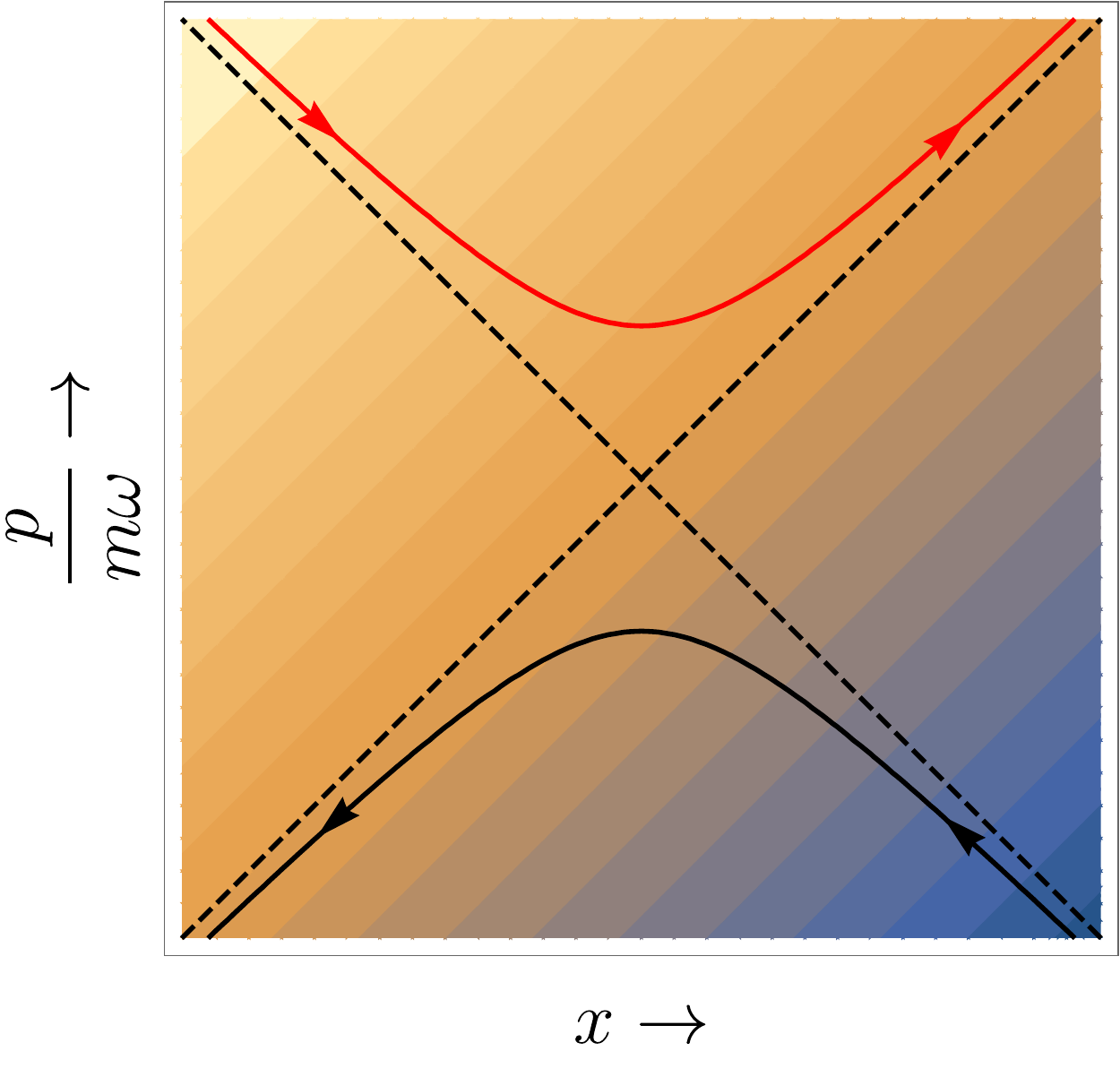}
		\caption{}
		\label{class_paths_E_positive}
	\end{subfigure}
	\begin{subfigure}[b]{0.4\textwidth}
		\includegraphics[width=1\textwidth]{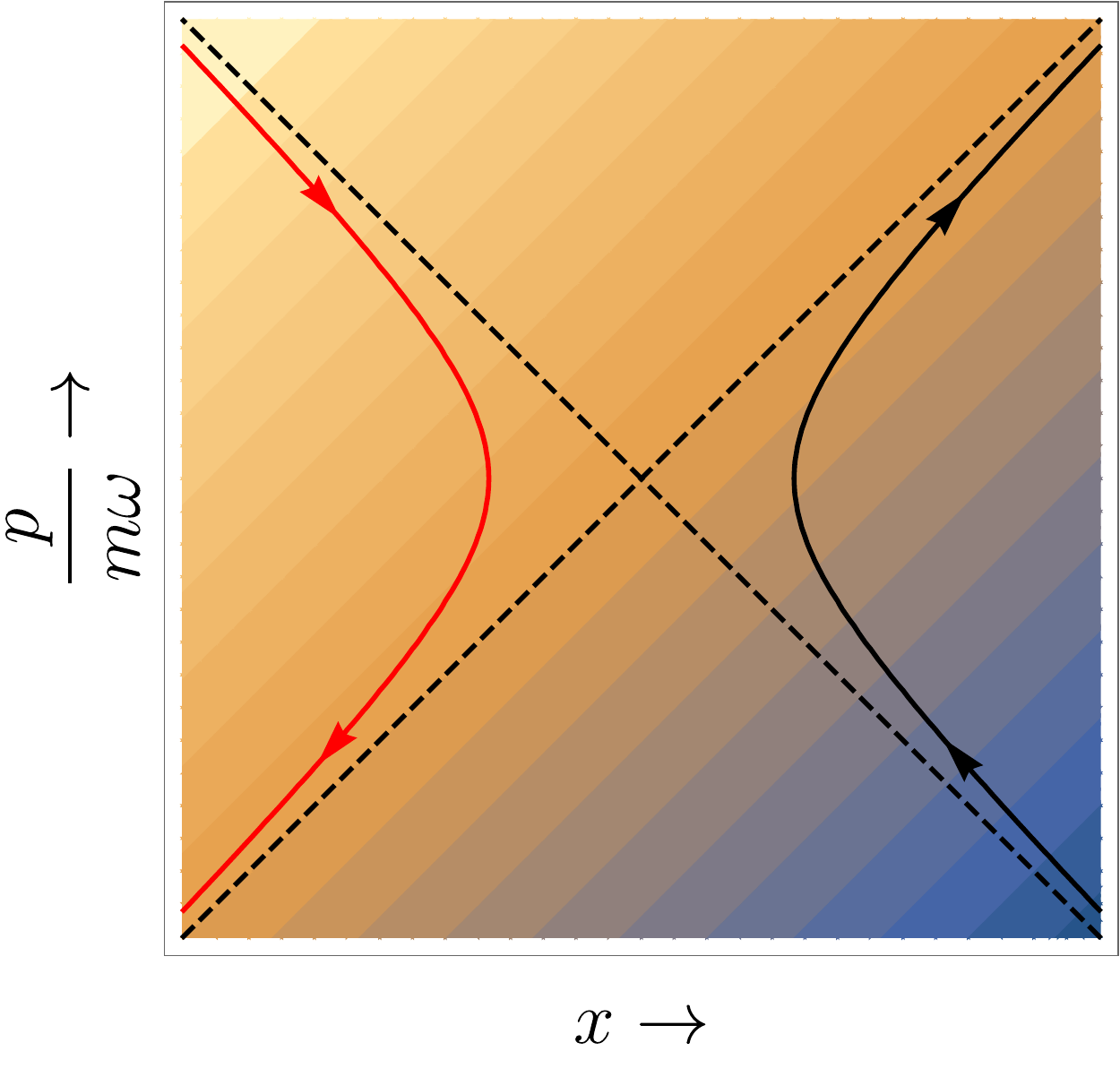}
		\caption{}
		\label{class_paths_E_negative}
	\end{subfigure}
	\caption{The classical trajectories of IHO in phase space: the cases of  positive and negative energies are shown in (a) and (b), respectively. The black continuous curves are solutions corresponding to particle incident towards the left from $x\rightarrow\infty$, while the red continuous curves represent particle incident towards the right from $x\rightarrow-\infty$. The arrows represent the positive sense of time. The lines of $p^2-m\omega^2x^2=0$ are shown in dashed black. The colour coded shadings represent contours of the function $p-m\omega x$; darker shades of blue represent more negative values and lighter shades of orange represent more positive values.}
	\label{class_paths}
\end{figure}

For better clarity of the problem, it is instructive to look at the classical trajectories in the phase space, which we have presented in \ref{class_paths}. The solutions to \ref{IHO_EOM} of our current interest have been shown in black continuous curves. The contours of the function $p-m\omega x$ are also represented in colour coded shadings. Notice that, \textit{at any instant}, the black curves satisfy:
\begin{align}\label{initial_cond_form1}
	p-m\omega x<0\quad;\quad\forall t\in\mathbb{R}
\end{align} 
Equivalently, we find that along the black curves we also have:
\begin{align}\label{initial_cond_form2}
	\frac{d}{dt}\left(p-m\omega x\right)>0\quad;\quad\forall t\in\mathbb{R}
\end{align}
Hence, the initial condition that unambiguously fixes the solution to be a left moving classical trajectory can be imposed as:
\begin{align}\label{initial_condition}
	\dot{x}(T_0)-\omega x(T_0)=\sigma_0<0,\quad\textrm{or equivalently}\quad \ddot{x}(T_0)-\omega \dot{x}(T_0)=\bar{\sigma}_0>0.
\end{align}
We shall be working with the former condition and, for convenience, write the constant $\sigma_0$ in the form $\sigma_0=-\omega u_0e^{-\omega T_0}$, where $u_0>0$. 

The solution to the equation of motion \ref{IHO_EOM} satisfying the boundary conditions \ref{initial_condition} and \ref{final_condition} turns out to be:
\begin{align}\label{class_sol_invsho}
x_{L}(t)=\frac{1}{2}u_0e^{-\omega t}-\frac{\epsilon}{m u_0 \omega^2}e^{\omega t}
\end{align} 
where,
\begin{align}
\epsilon=\frac{1}{2} m u_0 \omega ^2 e^{-T \omega } \left(u_0 e^{-T\omega }-2 x_1\right)
\end{align}
is the total energy and the subscript `$L$' indicates that the solution describes particle moving towards the left in the asymptotic past. It is interesting to see that the initial time $T_0$ does not appear explicitly in the solution $x_{L}(t)$, implying that we are at liberty to choose it at any value. The positive/negative energy solutions are characterised by $(e^{-\omega T}u_0-2x_1)$ being positive/negative, respectively. It is worth mentioning that the right moving solution $x_{R}(t)$ can be obtained by replacing $u_0\rightarrow-u_0$ in $x_{L}(t)$, to obtain:
\begin{align}\label{class_sol_invsho_2}
x_{R}(t)=-\frac{1}{2}u_0e^{-\omega t}+\frac{\epsilon}{m u_0 \omega^2}e^{\omega t}
\end{align} 
We shall now see how the boundary conditions \ref{initial_condition} and \ref{final_condition} can be naturally incorporated into the variational problem by adding an appropriate boundary term to the action.

The standard action for the IHO system is given by
\begin{align}
	\mathcal{S}_0[x]=\int_{T_0}^{T}\left(\frac{1}{2}m\dot{x}^2+\frac{1}{2}m\omega^2x^2\right)dt
\end{align} 
Let us now define a modified action $\mathcal{S}[x;T_0]$, by adding a boundary term $\mathcal{B}[x(T_0)]$ to the standard action, where:
\begin{align}
	\mathcal{B}[x(T_0);u_0]=\frac{m \omega}{2}\left[x(T_0)-u_0 e^{-\omega T_0}\right]^2 
\end{align}
The modified action then becomes:
\begin{align}
	\mathcal{S}[x;u_0]=\mathcal{S}_0[x]+\mathcal{B}[x(T_0)]
\end{align}
Therefore, the variation of $\mathcal{S}$ takes the form:
\begin{align}\label{vary_S}
	\delta\mathcal{S}=\int_{T_0}^{T}\left(-m\ddot{x}+m\omega^2x\right)\delta xdt+m\dot{x}(T)\delta x(T)+m\left[-\dot{x}(T_0)+\omega x(T_0)-\omega u_0e^{-\omega T_0}\right]\delta x(T_0)
\end{align}
When $x(T_0)$ is not fixed, the above equation shows that the modified action $\mathcal{S}$ is suited for the variational problem in which Dirichlet condition, say $x(T)=x_1$, is applied at $t=T$ and a Robin boundary condition $\dot{x}(T_0)-\omega x(T_0)+\omega u_0 e^{-\omega T_0}=0$ is applied at $t=T_0$. Note that addition of boundary term to fix the state is not new, for instance,  see\cite{Khlebnikov:1991th}, where a similar approch was utilised to generate coherent states.  In what follows, we shall find that $\mathcal{S}[x;T_0]$, by virtue of naturally imposing the requisite boundary condition appropriate to the scattering problem, makes the path integral formulation straightforward. 
\subsection{The scattering wave function as a path integral amplitude}\label{PI_IHO_main}
Let us consider the path integral amplitude $\Psi^{(L)}_{\epsilon}(x_1,T)$ that naturally follows from the action $\mathcal{S}[x;u_0]$:
\begin{align}\label{def_Psi_L}
	\Psi^{(L)}_{\epsilon}(x_1,T)\equiv \int^{x(T)=x_1} \mathcal{D}[x] e^{\frac{i}{\hbar}\mathcal{S}[x;u_0]}
\end{align}
where, once again, the superscript `$(L)$' indicates that the solution describes particle moving towards the left. Note that, while there is a fixed final boundary condition for the paths that are being summed over in \ref{def_Psi_L}, the initial value $x(T_0)$ can take all values in the real line. The above path integral can be explicitly evaluated to get:
\begin{align}\label{Phi_def}
	\Psi_{\epsilon}^{(L)}(x_1|T)=e^{-\frac{1}{2}\omega(T-T_0)}\exp\left(\frac{i}{\hbar}\mathcal{S}[x_L;u_0]\right)
\end{align}
where, $\mathcal{S}[x_L;u_0]$ is the modified action evaluated at the classical solution $x_{L}(t)$. The above wave function is not normalizable, but that is not surprising for a scattering solution of the Schr\"{o}dinger equation. Besides, the scattering amplitudes can be found out without fixing the normalization, so we shall not bother about the same. Another point worth mentioning is, that the fluctuation factor in \ref{Phi_def}, viz. $\mathcal{F}(T,T_0)\equiv e^{-\omega(T-T_0)/2}$ is simply one over square root of the determinant of IHO operator $(\partial_{t}^2-\omega^2)$. However, the relevant space of fluctuations, on which $(\partial_{t}^2-\omega^2)$ acts, are those that satisfy the boundary conditions $\delta x'(T_0)-\omega \delta x(T_0)=0$ and $\delta x(T)=0$, in view of \ref{initial_condition} and \ref{final_condition}. In contrast, while solving for the propagator of IHO, one considers space of fluctuations that satisfy $\delta x(T_0)=\delta x(T)=0$. In the case of propagator, Gelfand-Yaglom formula dictates that the determinant is proportional to $D(T)$, where $D(t)$ is the solution of IHO equation, satisfying the initial condition $D(T_0)=0$. In an analogous way, the determinant that leads to the fluctuation factor in \ref{Phi_def} is proportional to $\tilde{D}(T)$, where $\tilde{D}(t)$ is a solution of the IHO equation, satisfying the initial condition $\tilde{D}'(T_0)-\omega \tilde{D}(T_0)=0$. The solution $\tilde{D}(t)$ is easily found to be $\tilde{D}(t)\propto e^{\omega(t-T_0)}$, which leads to $\mathcal{F}(T,T_0)\propto[\tilde{D}(T)]^{-1/2}\propto e^{-\omega(T-T_0)/2}$.  

We shall now construct a stationary solution of the Schr\"{o}dinger equation, that corresponds to the scattering problem in inverted harmonic oscillator potential. A stationary solution of energy $\epsilon$, may be interpreted as describing a steady influx of incident particles, which are reflected and transmitted in the potential. If $\Psi^{(L)}_{\epsilon}(x_1|T)$ describes a particle $P_1$ incident on the potential barrier, then $\Psi^{(L)}_{\epsilon}(x_1|T-\tau)$ describes a particle incident on the potential barrier after a time $\tau$ that the particle $P_1$ was incident. Therefore, a stream of particles can be described by the following superposition:
\begin{align}
	\bar{\Psi}^{(L)}_{\epsilon}(x_1|T)=\int_{-\infty}^{\infty}f(\tau)\Psi^{(L)}_{\epsilon}(x_1|T-\tau)\, d(\omega\tau)
\end{align} 
where, $f(\tau)$ is some function. However, for the above superposition to describe a stationary state, the function $f(\tau)$ should have the form: $f(\tau)=f_0e^{-i\epsilon\tau}$, in which case the above integral, after a change of variable, reduces to:
\begin{align}
	\tilde{\Psi}^{(L)}_{\epsilon}(x_1|T)&=f_0e^{-i\epsilon T}\int_{-\infty}^{\infty}\Psi^{(L)}_{\epsilon}(x_1|\tau)e^{i\epsilon\tau} \,d(\omega\tau)\\\label{phi_def}
	&\equiv f_0e^{-i\epsilon T} \psi^{(L)}_{\epsilon}(x_1)
\end{align}
where, $\psi^{(L)}_{\epsilon}(x_1)$ is the solution to time-independent Schr\"{o}dinger equation with the fixed energy $\epsilon$. This implies that the wave function $\psi^{(L)}_{\epsilon}(x_1)$  has the following path integral representation:
\begin{align}\label{psi_L_PI}
	\psi^{(L)}_{\epsilon}(x_1)\propto\int_{-\infty}^{\infty}\int^{x_1} \exp\left[\frac{i}{\hbar}\mathcal{S}[x;u_0]+\frac{i}{\hbar}\epsilon(T-T_0)\right]\mathcal{D}[x]\, dT
\end{align}
Using \ref{Phi_def}, followed by a suitable variable change, we can show that the path integral amplitude $\psi^{(L)}_{\epsilon}(x_1)$ above is, in fact, the same standard scattering wave function in IHO that we have introduced in \ref{psi_L_first_appear}. We have delegated the details of this calculation to \ref{exact_psi_int}.

\subsection{Semi-classical scattering amplitudes from Picard-Lefschetz theory}

Let us first rewrite \ref{psi_L_PI} in the following form:
\begin{align}\label{phi_def_2}
\psi^{(L)}_{\epsilon}(x_1)=\mathcal{N} \int_{-\infty}^{\infty}e^{-\frac{\omega T}{2}}\exp\left[\frac{i}{\hbar}\tilde{\mathcal{S}}(T;\epsilon,x_1,u_0)\right]\,dT
\end{align}
where we have used \ref{Phi_def}, then absorbed all the constant factors into a single normalization coefficient $\mathcal{N}$ and defined
\begin{align}\label{S_epsilon}
i\tilde{\mathcal{S}}(T;\epsilon,x_1,u_0)&= \frac{i m \omega}{2}\left(x_1-u_0 e^{-\omega T}\right)^2-\frac{i m u_0^2 \omega }{4}e^{-2\omega T }+i\epsilon T.
\end{align}
In a similar fashion, one can also obtain the `right-moving' solution $\psi^{(R)}_{\epsilon}(x_1)$ by replacing $u_0\rightarrow-u_0$ in the right hand side of \ref{phi_def_2}, which yields:  
\begin{align}\label{phi_def_rightmov}
\psi^{(R)}_{\epsilon}(x_1)&= \mathcal{N} \int_{-\infty}^{\infty}e^{-\frac{\omega T}{2}}\exp\left[\frac{i}{\hbar}\tilde{\mathcal{S}}(T;\epsilon,x_1,-u_0)\right]\,dT
\end{align}
The above wave function describes a stream of particles incident towards the right from $x\rightarrow-\infty$ which, in turn, gets reflected and transmitted in the IHO potential. Note that the $T-$integrals in \ref{phi_def_2} and \ref{phi_def_rightmov} are not absolutely convergent, owing to the oscillatory exponential factors in the integrands. In real time based path integral approaches, one ought to be confronted with oscillatory integrals such as the above. In such cases, investigators are often intimidated by the fact that the relevant integral is not absolutely convergent and migrate to the Euclidean version of the problem to make progress. Although the $T$-integral in \ref{phi_def_2} is not absolutely convergent, as it turns out, it is conditionally convergent. In this section, instead of resorting to the Euclidean approach, we shall transform the right hand side of \ref{phi_def_2} into an absolutely convergent integral by utilizing Picard-Lefschetz theory. Further, we shall derive the semi-classical scattering amplitudes by performing the saddle point approximation. A similar analysis can also be done for \ref{phi_def_rightmov} as well, but we shall not elaborate the same here. It is worth mentioning that the application of real-time path integral formalism and Picard-Lefshetz theory for studying tunnelling is well explored in several other contexts (for instance, see \cite{Ai:2019fri,Tanizaki:2014xba,Matsui:2021oio}). However, to our knowledge, the particular analysis presented in this section has not appeared at this level of details elsewhere.

Following Picard-Lefschetz theory, we start by interpreting $\tilde{\mathcal{S}}(T;\epsilon,x_1,u_0)$ as a holomorphic function in the complex $T$-plane. This leads to the following expressions for the real and imaginary parts of $\tilde{\mathcal{S}}(T;\epsilon,x_1,u_0)$:
\begin{align}
	i\tilde{\mathcal{S}}(T;\epsilon,x_1,u_0)&=\left[-m x_1 e^{-X} u_0 \omega  \sin (Y)+\frac{1}{4} m e^{-2 X} u_0^2 \omega  \sin (2 Y)-\frac{Y \epsilon	}{\omega }\right]\\\nonumber
	&+i\left[\frac{1}{2} m x_1^2 \omega -m x_1 e^{-X} u_0	\omega  \cos (Y)+\frac{1}{4} m e^{-2 X} u_0^2	\omega  \cos (2 Y)+\frac{X \epsilon }{\omega }\right]
\end{align}
where, $\omega T\equiv (X+iY)$. The saddle points of $\tilde{\mathcal{S}}(T;\epsilon,x_1,u_0)$, which are point at which $\partial_{T}\tilde{\mathcal{S}}(T;\epsilon,x_1,u_0)=0$, turns out to be:
\begin{align}
	\omega T_{n,\pm}=-\log \left(\frac{x_1}{u_0}\pm\sqrt{\frac{2 \epsilon}{m\omega^2u_0^2}+\frac{x_1^2}{u_0^2}}\right)+2\pi i n\quad;\quad n\in\mathbb{Z}
\end{align} 
For oscillatory integrals such as \ref{phi_def_2}, Picard-Lefschetz theory offers a prescription to determine the saddle points that are relevant. We shall not attempt to make this article self-contained on the subject of Picard-Lefschetz theory; for a useful discussion, the reader may consult \cite{Feldbrugge:2017kzv}. The key result that we require here may be summarized as follows: the relevant saddle points are those, whose steepest ascent curves intersect the original integration contour an odd number of times. There is a caveat, though: In our specific problem, the steepest descent/ascent contours are degenerate, which means that the steepest ascent/descent curve originating from certain saddle points end up intersecting other saddle points. This is due to the periodicity of $\textrm{Im}[i\tilde{\mathcal{S}}(T;\epsilon,x_1,u_0)]$ along the Euclidean direction. When the degeneracies of the steepest ascent/descent contours can be traced to a certain symmetry, a standard strategy is to add an infinitesimal term to $i\tilde{\mathcal{S}}(T;\epsilon,x_1,u_0)$ that breaks this symmetry, which, in turn, breaks the degeneracy of the ascent/descent contours. In the present case, the most convenient way to achieve this is to add a small positive imaginary component to the energy\footnote{One could have, instead, added an infinitesimal negative imaginary component to the energy. The analysis follows almost exactly the same way, with only minor departures. The final results, however, remain exactly the same.}: $\epsilon\rightarrow\epsilon+i\delta$, where $\delta>0$. The real and imaginary parts of $i\tilde{\mathcal{S}}(T;\epsilon,x_1,u_0)$ now becomes:
\begin{align}
	{\rm Re}[i\tilde{\mathcal{S}}(T;\epsilon,x_1,u_0)]&=-m x_1 e^{-X} u_0 \omega  \sin (Y)+\frac{1}{4} m e^{-2 X} u_0^2 \omega  \sin (2 Y)-\frac{\delta	X}{\omega }-\frac{Y \epsilon }{\omega }\\
	{\rm Im}[i\tilde{\mathcal{S}}(T;\epsilon,x_1,u_0)]&=\frac{1}{2} m x_1^2\omega -m x_1 e^{-X} u_0 \omega  \cos (Y)+\frac{1}{4}	m e^{-2 X} u_0^2 \omega  \cos (2 Y)+\frac{X	\epsilon }{\omega }-\frac{\delta  Y}{\omega }
\end{align} 
Clearly, ${\rm Im}[\tilde{\mathcal{S}}(T;\epsilon,x_1,u_0)]$ is not a periodic function of $Y$ now. The saddle points, on the other hand, become:
\begin{align} 
	\omega T^{(\delta)}_{n,\pm}=\omega T_{n,\pm}+\mathcal{O}(\delta)
\end{align} 

\subsubsection{A manual to interpret the figures }\label{scheme_plot}
To facilitate the application of Picard-Lefschetz theory, we have introduced a simple scheme for visualizing the saddle points and their steepest descent/ascent contours in the complex $T-$plane. Before going into the details of the calculations, we present a summary of this scheme: (i) the saddle points are represented by small solid circles of black and red colors, denoting the saddle points $T^{(\delta)}_{n,+}$ and $T^{(\delta)}_{n,-}$, respectively, (ii) steepest ascent/descent contours emanating from saddle points $T^{(\delta)}_{n,+}$ are represented by black curves, while that from $T^{(\delta)}_{n,-}$ are represented by red curves, (iii) in addition, the steepest ascent/descent contours from relevant saddle points are denoted by continuous curves, while that from other saddle points are denoted by dotted curves, (iv) to better visualise the ascent/descent directions, contours of $\textrm{Re}[i\tilde{\mathcal{S}}(T;\epsilon,x_1,u_0)]$ are also shown in colour coded shadings; lighter shades of orange correspond to more positive values, while darker shades of blue correspond to more negative values, (v) the original, real time contour in \ref{phi_def_2} is denoted by horizontal, continuous green line and (vi) the complex contours, to which the real time contour should be deformed, are denoted by dashed green curves. 

When we apply the above scheme and make the relevant plots in the complex $T-$plane, we see that the cases of positive and negative energy are visibly different. This is related to the fact that nature of classical trajectories in an IHO potential are quit different for the cases of negative and positive energies, as can be seen in \ref{class_paths}. Specifically, while there are real turning points for the former case, the same are absent in case of the latter. The presence or absence of real turning points acquire interesting interpretations in terms of the nature and relevance of the saddle points of $\tilde{\mathcal{S}}(T;\epsilon,x_1,u_0)$. Hence, it is instructive to consider the positive and negative energy cases separately. 
\subsubsection{Case 1: Over-the-barrier reflection ($\epsilon>0$)}
One finds that the saddle point $T_{0,+}$ is always real for all values of $x_1$. In fact, it corresponds to the classical solution in \ref{class_sol_invsho}, as can be seen by writing:
\begin{align}
	x_1=\frac{1}{2}u_0e^{-\omega T_{0,+}}-\frac{\epsilon}{m u_0 \omega^2}e^{\omega T_{0,+}},
\end{align} 
which is not surprising. Hence, one would expect that $\omega T_{0,+}$ will always contribute to the saddle point approximation of $\phi_{\epsilon}(x_1)$. However, as we shall shortly see, depending on whether $x_1$ is negative or positive, one gets a contribution from another saddle point as well. This is nothing but manifestation of the well known Stokes phenomenon. In view of this, let us study the wave function $\psi^{(L)}_{\epsilon}(x_1)$ in the two region $x_1<0$ and $x_1>0$, separately. 
\subsubsection*{The region $x_1<0$ } 
\begin{figure}[h!]
	\begin{subfigure}{.5\linewidth}
		\centering
		\includegraphics[scale=.6]{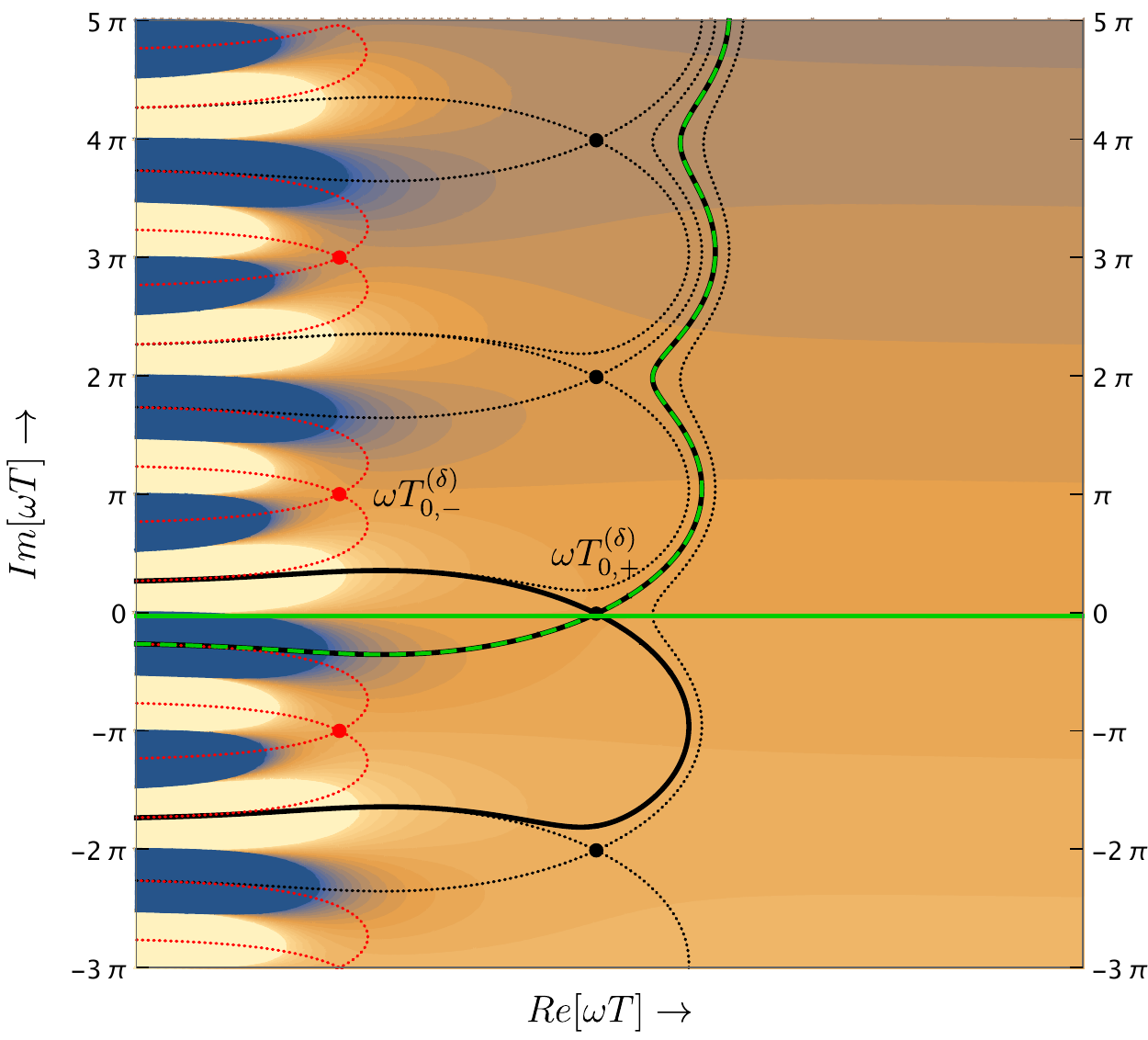}
		\caption{}
		\label{contours_Epos_xneg}
	\end{subfigure}
	\begin{subfigure}{.5\linewidth}
		\centering
		\includegraphics[scale=.6]{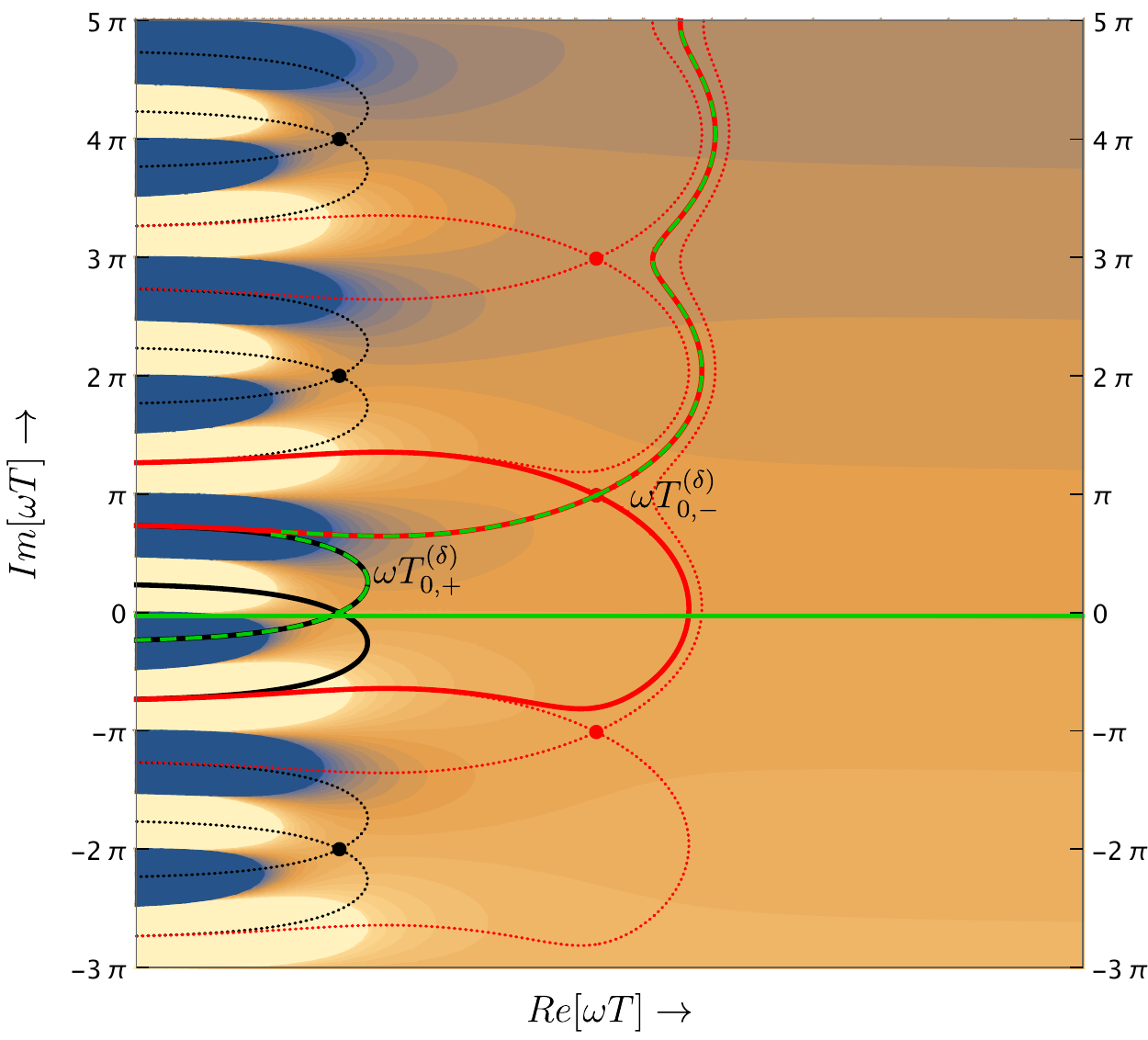}
		\caption{}
		\label{contours_Epos_xpos}
	\end{subfigure}
\vspace{.5 cm}

\hspace{.25\linewidth}
\begin{subfigure}{.5\linewidth}
	\centering
	\includegraphics[scale=.6]{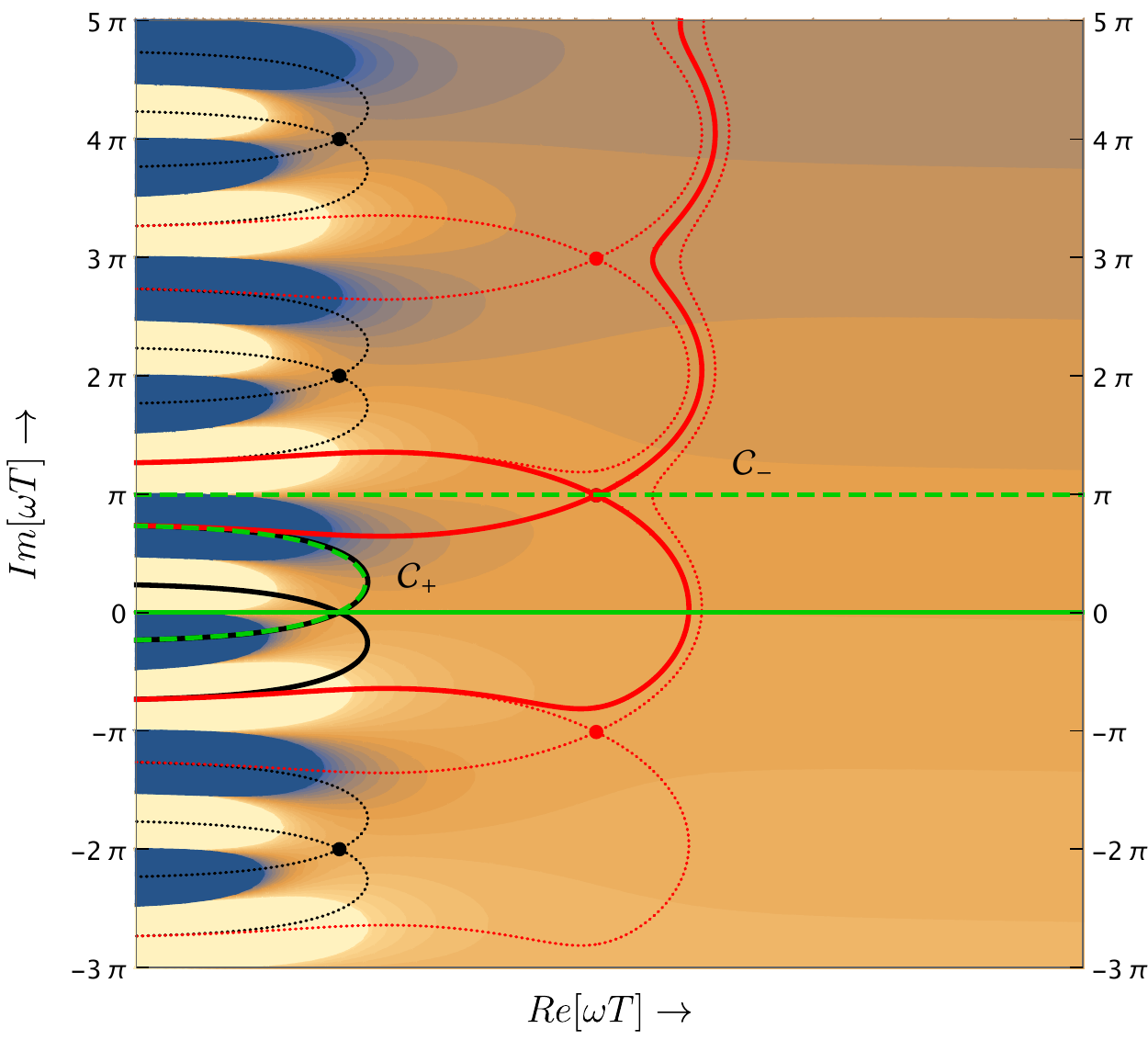}
	\caption{}
	\label{contours_Epos_xpos_exact}
\end{subfigure}
\caption{Plots of saddle points and steepest descent/ascent lines when $\epsilon>0$, using the convention described in \ref{scheme_plot}. (a) Region $x_1<0$. The dashed green curve is $\mathcal{T}_{a}$. (b) Region $x_1>0$. While, the contour $\mathcal{T}_{b,+}$ is represented by the dashed green curve passing through $\omega T^{(\delta)}_{0,+}$,  the contour $\mathcal{T}_{b,-}$ is represented by the dashed green curve passing through $\omega T^{(\delta)}_{0,-}$. (c) Region $x_1>0$. The horizontal dashed contour $\mathcal{C}_{-}$ is required for exact computations described in \ref{alpha_derivation}.}
\label{contour_Epos}
\end{figure}
The relevant plot for this case is given in \ref{contours_Epos_xneg}. Note that the steepest ascent contour from only $ T^{(\delta)}_{0,+}$ is intersecting the original integration contour. Using Picard-Lefschetz theory, we conclude that the only relevant saddle point is $T^{(\delta)}_{0,+}$. Hence, the original, real time contour can be smoothly into the green dashed contour in \ref{contours_Epos_xneg}, which we shall denote as $\mathcal{T}_{a}$. Therefore, when $x_1<0$, the wave function $\psi^{(L)}_{\epsilon}(x_1)$ can be written as  
\begin{align}\label{psi_Epos_left}
\psi^{(L)}_{\epsilon}(x_1)= \mathcal{N} \int_{\mathcal{T}_{a}}e^{-\frac{\omega T}{2}}\exp\left[\frac{i}{\hbar}\tilde{\mathcal{S}}(T;\epsilon,x_1,u_0)\right]\,dT,\quad;\quad x_1<0
\end{align}  
Since $\mathcal{T}_{a}$ is defined along the steepest descent contour of the relevant saddle point, the above integral is absolutely convergent. The semi-classical limit of $\psi_{\epsilon}^{(L)}(x_1)$ can be obtained from the saddle point approximation of the above integral, which yields:
\begin{align}\label{transmitted_Epos}
	\psi^{(L)}_{\epsilon}(x_1)\propto \left.\exp\left[\frac{i\tilde{\mathcal{S}}(T;\epsilon,x_1,u_0)}{h}-\frac{1}{2}\omega T\right]\right|_{T^{(\delta)}_{0,+}}=\mathcal{N}_1 \frac{e^{-\frac{i}{\hbar}\mathcal{S}_{+}(x_1)}}{\sqrt{\left|\partial_{x_1}\mathcal{S}_{+}(x_1)\right|}}
\end{align}
where, $\mathcal{N}_1$ is a constant pre-factor and
\begin{align}
	\mathcal{S}_{+}(x_1)=\frac{\epsilon  }{\omega }\log \left(\sqrt{\frac{2 \epsilon }{m u_0^2\omega^2}+\frac{x_1^2}{u_0^2}}+\frac{x_1}{u_0}\right)+\frac{1}{2} x_1 \sqrt{m \left(m x_1^2 \omega ^2+2\epsilon \right)}
\end{align}
Note that semi-classical limit of $\psi^{(L)}_{\epsilon}(x_1)$ has the form of a purely left moving wave. This is expected, since, the only relevant saddle point is $T_{0,+}^{(\delta)}$, which corresponds to the left moving classical solution $x_{L}(t)$. 
\subsubsection*{The region $x_1>0$}
The plots for this case are given in \ref{contours_Epos_xpos}. We find that steepest ascent contours from both $T^{(\delta)}_{0,+}$ and $T^{(\delta)}_{0,-}$ intersect the original contour once. However, unlike $T^{(\delta)}_{0,+}$, the saddle point $T^{(\delta)}_{0,-}$ is complex.
\begin{align}
	\omega T^{(\delta)}_{0,-}=-\log \left(\frac{x_1}{u_0}-\sqrt{\frac{2 \epsilon}{m\omega^2u_0^2}+\frac{x_1^2}{u_0^2}}-i0^{+}\right)=-\log \left(\sqrt{\frac{2 \epsilon}{m\omega^2u_0^2}+\frac{x_1^2}{u_0^2}}-\frac{x_1}{u_0}-i0^{+}\right)+i\pi
\end{align}   
where, $0^{+}$ denotes the infinitesimal correction introduced by $\delta$. The real line contour can now be deformed into one that has two parts, denoted henceforth by $\mathcal{T}_{b,+}$ and $\mathcal{T}_{b,-}$, which are curves along steepest descent contours of $T^{(\delta)}_{0,+}$ and $T^{(\delta)}_{0,-}$, respectively. The deformed contour $\mathcal{T}_{b}=\mathcal{T}_{b,+}+\mathcal{T}_{b,-}$ is represented by the dashed green curve in \ref{contours_Epos_xpos}. The wave function $\psi^{(L)}_{\epsilon}(x_1)$ can now be written as:
\begin{align}\label{psi_Epos_right}
\psi^{(L)}_{\epsilon}(x_1)&= \mathcal{N} \int_{\mathcal{T}_{b}}e^{-\frac{\omega T}{2}}\exp\left[\frac{i}{\hbar}\tilde{\mathcal{S}}(T;\epsilon,x_1,u_0)\right]\,dT\\\label{E_pos_int_parts}
&=\mathcal{N} \int_{\mathcal{T}_{b,+}}e^{-\frac{\omega T}{2}}\exp\left[\frac{i}{\hbar}\tilde{\mathcal{S}}(T;\epsilon,x_1,u_0)\right]\,dT+\mathcal{N} \int_{\mathcal{T}_{b,-}}e^{-\frac{\omega T}{2}}\exp\left[\frac{i}{\hbar}\tilde{\mathcal{S}}(T;\epsilon,x_1,u_0)\right]\,dT
\end{align} 
Note that the above integrals are absolutely convergent. The semi-classical limit of $\psi^{(L)}_{\epsilon}(x_1)$, in the region $x_1>0$, can be obtained by performing the saddle point approximation to both parts of \ref{E_pos_int_parts}. The contribution from $T^{(\delta)}_{0,-}$ gives a reflected wave of the form:
\begin{align}
	\mathcal{N} \int_{\mathcal{T}_{b,-}}e^{-\frac{\omega T}{2}}\exp\left[\frac{i}{\hbar}\tilde{\mathcal{S}}(T;\epsilon,x_1,u_0)\right]\,dT&\propto \left.\exp\left[\frac{\tilde{\mathcal{S}}(T;\epsilon,x_1,u_0)}{h}+\frac{1}{2}\omega T\right]\right|_{T^{(\delta)}_{0,-}}\\
	&=e^{-\frac{\pi\epsilon}{\hbar\omega}}e^{-i\frac{\pi}{2}}\mathcal{N}_1 \frac{e^{\frac{i}{\hbar}\mathcal{S}_{+}(x_1)}}{\sqrt{\left|\partial_{x_1}\mathcal{S}_{+}(x_1)\right|}}\\\label{reflected_Epos}
	&\approx e^{-\frac{\pi\epsilon}{\hbar\omega}}e^{-i\frac{\pi}{2}}\psi^{(R)}_{\epsilon}(x_1)
\end{align}
where, $\psi^{(R)}_{\epsilon}(x_1)$ is the `right-moving' wave function introduced in \ref{phi_def_rightmov}.
Comparing \ref{transmitted_Epos} and \ref{reflected_Epos}, we conclude that the ratio of semi-classical probability for transmission $\mathcal{P}_{t}$, to that for reflection $\mathcal{P}_{r}$, must be given by:
\begin{align}
	\frac{\mathcal{P}_{r}}{\mathcal{P}_{t}}=e^{-\frac{2\pi\epsilon}{\hbar\omega}}
\end{align}
from which one can easily compute $\mathcal{P}_{t}$ and $\mathcal{P}_{r}$, using the identity $\mathcal{P}_{t}+\mathcal{P}_{r}=1$.

\subsubsection{Case 2: Quantum tunnelling ($\epsilon<0$)}
\begin{figure}[h!]
	\begin{subfigure}{0.5\linewidth}
		\centering
		\includegraphics[scale=.6]{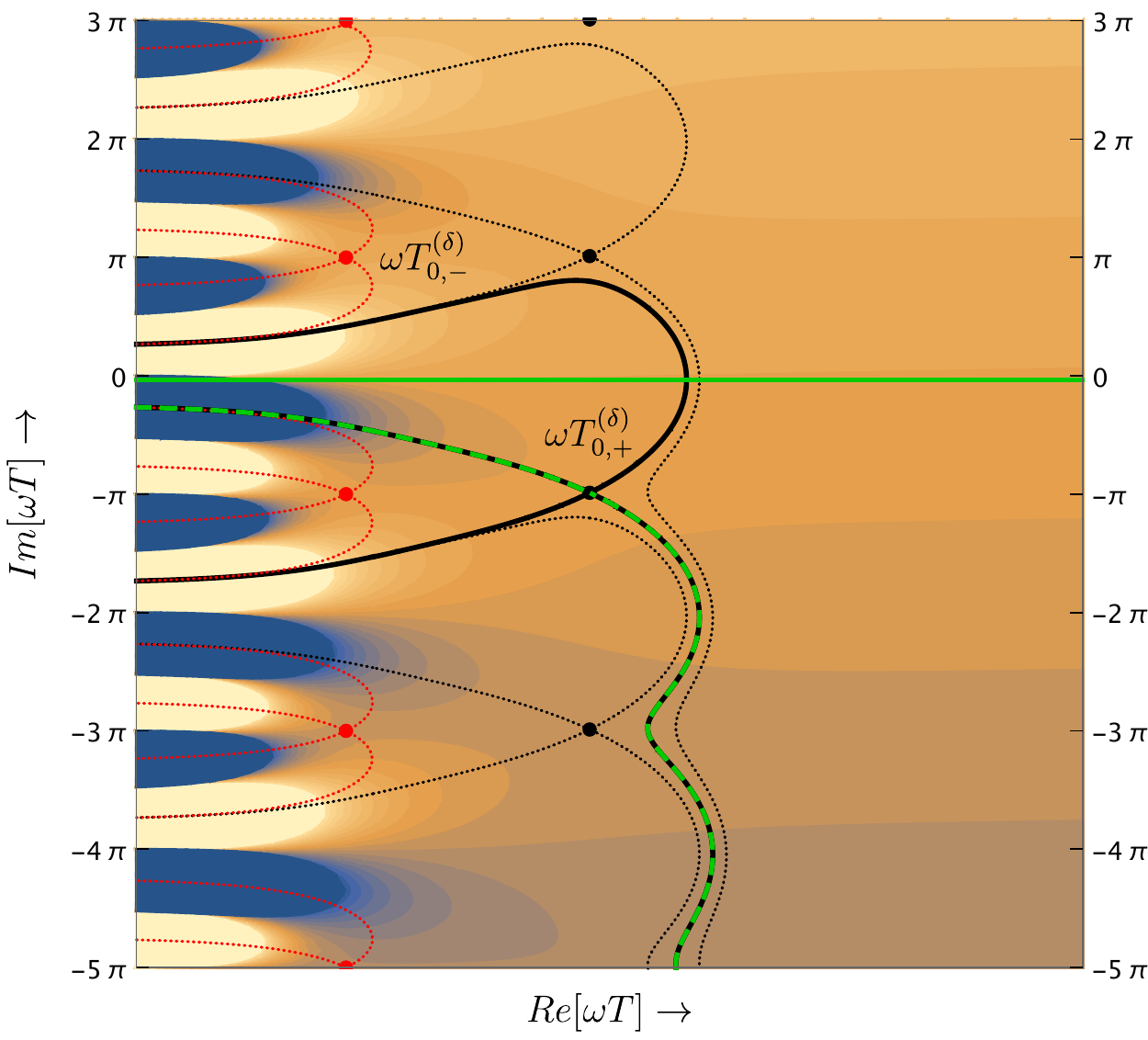}
		\caption{}
		\label{contours_Eneg_left}
	\end{subfigure}
	\begin{subfigure}{0.5\linewidth}
		\centering
		\includegraphics[scale=.6]{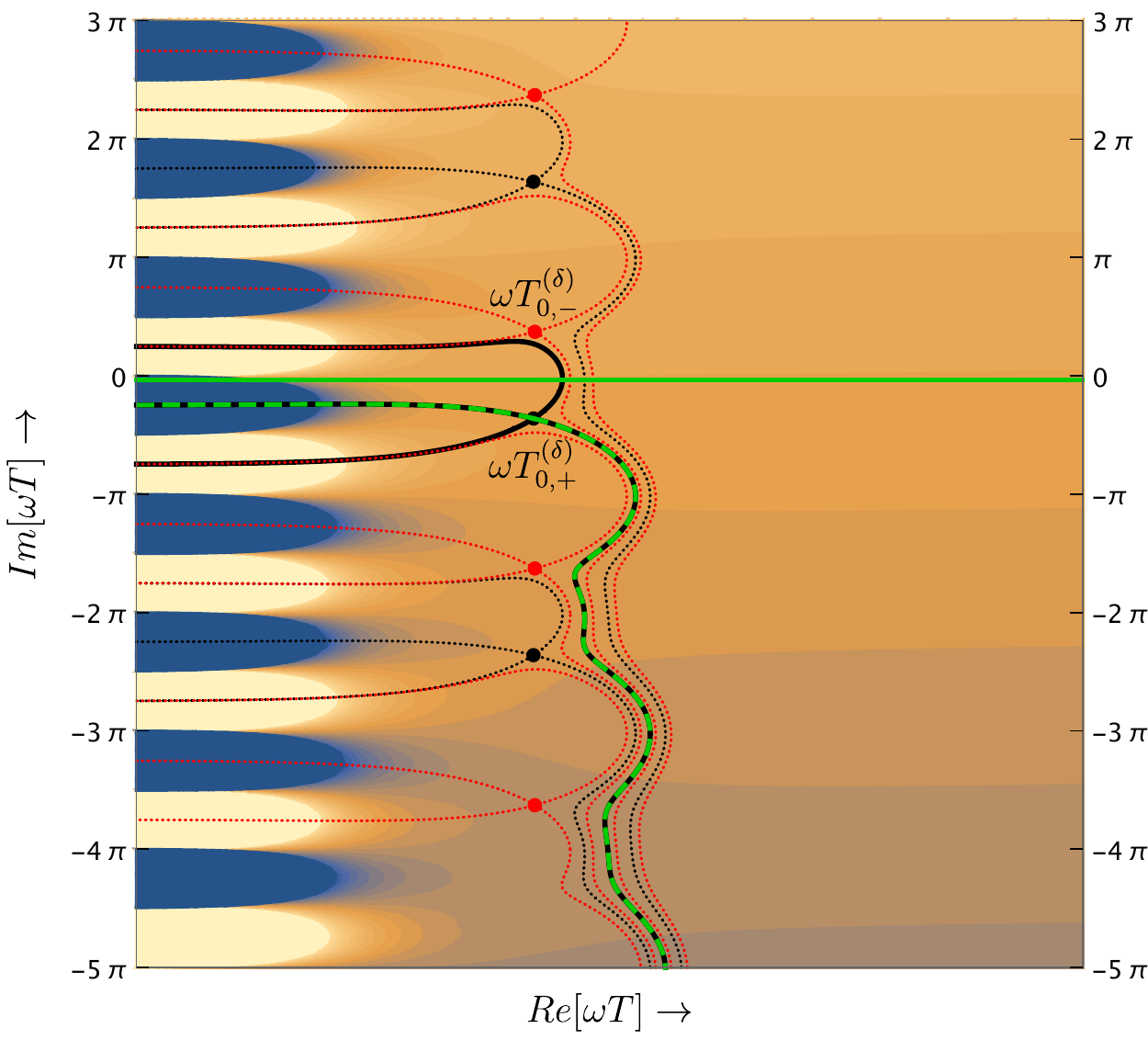}
		\caption{}
		\label{contours_Eneg_mid}
	\end{subfigure}
\vspace{.5 cm}

\hspace{.25\linewidth}
	\begin{subfigure}{0.5\linewidth}
		\centering
		\includegraphics[scale=.6]{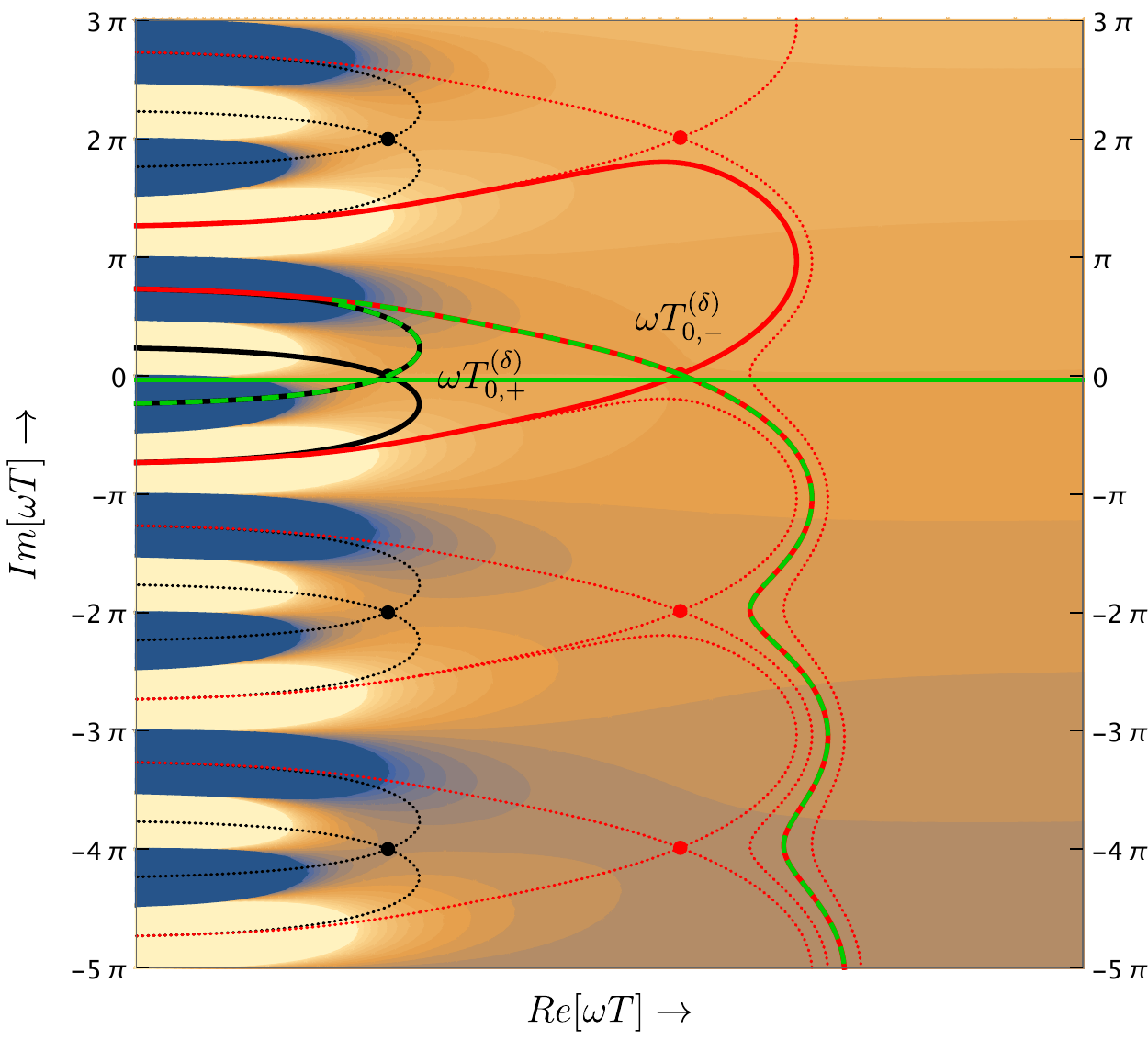}
		\caption{}
		\label{contours_Eneg_right}
	\end{subfigure}
	\caption{Plots of saddle points and steepest descent/ascent lines when $\epsilon<0$, using the convention described in \ref{scheme_plot} (Plots are generated for values $\omega=m=u_0=1$, $\epsilon=-3$ and $\delta=0.1$). (a) The Left region (Plots are for $x_1=-6$). The green dashed contour represents $\Upsilon_{a}$ . (b) The Middle region (Plots are for $x_1=1$). The green dashed contour represents $\Upsilon_{b}$ . (c) The Right region (Plots are for $x_1=6$). While the greed dashed curve passing through $\omega T^{(\delta)}_{0,+}$ represents $\Upsilon_{c,+}$, the one through $\omega T^{(\delta)}_{0,-}$ represents $\Upsilon_{c,-}$.}
	\label{contour_Eneg}
\end{figure}

In this case, the saddle points are given by:
\begin{align}
\omega T_{n,\pm}=-\log \left(\frac{x_1}{u_0}\pm\sqrt{\frac{x_1^2}{u_0^2}-\frac{2 |\epsilon|}{m\omega^2u_0^2}}\right)+2\pi i n\quad;\quad n\in\mathbb{Z}
\end{align}
The nature of saddle points is qualitatively different from the over-the-barrier reflection case. In particular, real saddle points exist only for $x_1>\sqrt{(2|\epsilon|/m\omega^2)}$, which corresponds to the classically allowed region. On the other hand, for $x_1<\sqrt{(2|\epsilon|/m\omega^2)}$, all the saddle points are complex. We shall now see how to determine the relevant saddle points that contribute to $\psi^{(L)}_{\epsilon}(x_1)$. For this purpose, it is convenient to define three distinct regions of space, namely-- Left, Middle and Right, as follows:
\begin{align}
	\textrm{Left}: -\infty<x_1<-\sqrt{\frac{2|\epsilon|}{m\omega^2}};&&
	\textrm{Middle}: -\sqrt{\frac{2|\epsilon|}{m\omega^2}}<x_1<\sqrt{\frac{2|\epsilon|}{m\omega^2}};&&
	\textrm{Right}: \sqrt{\frac{2|\epsilon|}{m\omega^2}}<x_1<\infty
\end{align} 
The steepest ascent/descent contours for this case are presented in \ref{contour_Eneg}. As we have already remarked, the relevant saddle points are the ones whose steepest ascent contours intersect the original real time contour an odd number of times.  

\subsubsection*{The Left region}
From \ref{contours_Eneg_left}, we see that the only relevant saddle point is $T^{(\delta)}_{0,+}$, which is complex. To determine the real and imaginary parts of $T^{(\delta)}_{0,+}$, we write: 
\begin{align}
\omega T^{(\delta)}_{0,+}=-\log \left(-\frac{|x_1|}{u_0}+\sqrt{\frac{x_1^2}{u_0^2}-\frac{2 |\epsilon|}{m\omega^2u_0^2}}+i0^{+}\right)=-\log \left(\sqrt{\frac{x_1^2}{u_0^2}-\frac{2 |\epsilon|}{m\omega^2u_0^2}}-\frac{|x_1|}{u_0}-i0^{+}\right)-i\pi
\end{align}
which is consistent with \ref{contours_Eneg_left}. Hence, an absolutely convergent integral representation for $\psi_{\epsilon}^{(L)}(x_1)$ can be obtained by deforming the real time contour to that represented by the dashed green curve in \ref{contours_Eneg_left}, that move along steepest descent contours of $T^{(\delta)}_{+,0}$. We shall denote this contour by $\Upsilon_{a}$, so that:
\begin{align}
	\psi^{(L)}_{\epsilon}(x_1)=\mathcal{N} \int_{\Upsilon_{a}}e^{-\frac{\omega T}{2}}\exp\left[\frac{i}{\hbar}\tilde{\mathcal{S}}(T;\epsilon,x_1,u_0)\right]\,dT
\end{align}
The semi-classical limit of $\psi^{(L)}_{\epsilon}(x_1)$, in the Left region, can be found by the saddle point approximation of the above integral, which yields:
\begin{align}\label{transmitted_Eneg}
\psi^{(L)}_{\epsilon}(x_1)\propto \left.\exp\left[\frac{i\tilde{\mathcal{S}}(T;\epsilon,x_1,u_0)}{h}-\frac{1}{2}\omega T\right]\right|_{T^{(\delta)}_{0,+}}=e^{-\frac{\pi|\epsilon|}{\hbar\omega}}e^{\frac{i\pi}{2}}\mathcal{N}_1 \frac{e^{-\frac{i}{\hbar}\mathcal{S}_{-}(x_1)}}{\sqrt{\left|\partial_{x_1}\mathcal{S}_{-}(x_1)\right|}}
\end{align}
where,
\begin{align}
\mathcal{S}_{-}(x_1)=\frac{\epsilon  }{\omega }\log \left(\sqrt{-\frac{2 |\epsilon| }{m u_0^2\omega^2}+\frac{x_1^2}{u_0^2}}+\frac{x_1}{u_0}\right)+\frac{1}{2} x_1 \sqrt{m \left(m x_1^2 \omega ^2-2|\epsilon| \right)}
\end{align}
\subsubsection*{The Middle region}
From \ref{contours_Eneg_mid}, we see that the only relevant saddle point is $T^{(\delta)}_{0,+}$ in this case as well. Hence, an absolutely convergent integral representation for $\psi_{\epsilon}^{(L)}(x_1)$ can be obtained by deforming the real time contour to that represented by the dashed green curve in \ref{contours_Eneg_mid}, which we shall denote by $\Upsilon_{b}$.
\begin{align}
\psi^{(L)}_{\epsilon}(x_1)=\mathcal{N} \int_{\Upsilon_{b}}e^{-\frac{\omega T}{2}}\exp\left[\frac{i}{\hbar}\tilde{\mathcal{S}}(T;\epsilon,x_1,u_0)\right]\,dT
\end{align}
Hence, the semi-classical limit of $\psi_{\epsilon}(x_1)$ takes the form:
\begin{align}\label{tunnel_Eneg}
\psi^{(L)}_{\epsilon}(x_1)\propto \left.\exp\left[\frac{i\tilde{\mathcal{S}}(T;\epsilon,x_1,u_0)}{h}-\frac{1}{2}\omega T\right]\right|_{T^{(\delta)}_{0,+}}=e^{-\frac{\pi|\epsilon|}{\hbar\omega}}e^{\frac{i\pi}{2}}\mathcal{N}_1 \frac{e^{\frac{\mathcal{S}_{E}(x_1)}{\hbar}}}{\sqrt{\left|\partial_{x_1}\mathcal{S}_{E}(x_1)\right|}}
\end{align}
where,
\begin{align}
\mathcal{S}_{E}(x_1)=-\frac{i\epsilon  }{2\omega }\log\left(\frac{2|\epsilon|}{m\omega^2u_0^2}\right)+\cos^{-1}\left(\sqrt{\frac{m\omega^2}{2|\epsilon|}}x_1\right)+\frac{x_1}{2}\sqrt{m \left(2|\epsilon|-m x_1^2 \omega ^2\right)}
\end{align}
The decaying nature of $\psi^{(L)}_{\epsilon}(x_1)$, known from standard WKB analysis, is reproduced in our approach.
\subsubsection*{The Right region}
From \ref{contours_Eneg_right}, we see that the relevant saddle points in this case are $T^{(\delta)}_{0,+}$ and $T^{(\delta)}_{0,-}$, respectively, corresponding to the incident and reflected waves. The real line contour can now be deformed into one that has two parts, denoted by $\Upsilon_{c,+}$ and $\Upsilon_{c,-}$, which are curves along steepest descent contours of $T^{(\delta)}_{0,+}$ and $T^{(\delta)}_{0,-}$, respectively. Hence, we obtain the following absolutely convergent integral representation for $\psi^{(L)}_{\epsilon}(x_1)$, for $x_1$ in the Right region:
\begin{align}
\psi^{(L)}_{\epsilon}(x_1)&=\psi^{i}_{\epsilon}(x_1)+\psi^{r}_{\epsilon}(x_1)
\end{align}
where, the incident and reflected parts $\psi^{i}_{\epsilon}(x_1)$ and $\psi^{r}_{\epsilon}(x_1)$ are, respectively, given by
\begin{align}\label{phi_plus_part_2}
\psi^{i}_{\epsilon}(x_1)&=\mathcal{N} \int_{\Upsilon_{c,+}}e^{-\frac{\omega T}{2}}\exp\left[\frac{i}{\hbar}\tilde{\mathcal{S}}(T;\epsilon,x_1,u_0)\right]\,dT\\\label{phi_minus_part_2}
\psi^{r}_{\epsilon}(x_1)&=\mathcal{N} \int_{\Upsilon_{c,-}}e^{-\frac{\omega T}{2}}\exp\left[\frac{i}{\hbar}\tilde{\mathcal{S}}(T;\epsilon,x_1,u_0)\right]\,dT
\end{align}
The reflected part, in particular, has the following semi-classical limit:
\begin{align}\label{reflec_eneg}
	\psi^{r}_{\epsilon}(x_1)\propto \left.\exp\left[\frac{i\tilde{\mathcal{S}}(T;\epsilon,x_1,u_0)}{h}-\frac{1}{2}\omega T\right]\right|_{T^{(\delta)}_{0,-}}=\mathcal{N}_1\frac{e^{\frac{i}{\hbar}\mathcal{S}_{-}(x_1)}}{\sqrt{\left|\partial_{x_1}\mathcal{S}_{-}(x_1)\right|}}
\end{align}
By comparing \ref{reflec_eneg} and \ref{tunnel_Eneg}, the ratio of transmission and reflection probabilities can be evaluated to get:
\begin{align}
	\frac{\mathcal{P}_t}{\mathcal{P}_r}=e^{-\frac{2\pi|\epsilon|}{\hbar\omega}}
\end{align}
from which, one can easily derive $\mathcal{P}_t$ and $\mathcal{P}_r$, using $\mathcal{P}_t+\mathcal{P}_r=1$.
\subsubsection{The emergence of instantons}\label{instanton}
Recall, that the paths contributing to $\psi^{(L)}_{\epsilon}$, by definition, satisfy the boundary condition:
\begin{align}
	x(T)=x_1
\end{align}
The saddle points $T_{n,\pm}$ correspond to those classical trajectories for which the total energy is $\epsilon$. When a relevant saddle point is complex, it means that in order to meet the above boundary condition, the corresponding classical trajectory has to evolve along imaginary time direction. To illustrate our point, we shall now consider only the positive energy case, but a similar argument can be easily extended to negative energy as well. In the case of positive energy, one of the relevant saddle points, namely, $T_{0,-}$, is complex when $x_1$ is in the region $x_1<0$. The real part of this saddle point satisfies the condition:
\begin{align}
x_{L}( {\rm Re}[T_{0,-}])=-x_1
\end{align}  
Now, one can imagine the complex time $T_{0,-}= {\rm Re}[T_{0,-}]+i\pi/\omega$ as describing the time evolution of $x_{L}(t)$ along the real time direction until $t= {\rm Re}[T_{0,-}]$, followed by an evolution along imaginary time for a duration $i\pi/\omega$. Defining the Euclidean time $t_{E}$ through $t={\rm Re}[T_{0,-}]+it_{E}$, we get the imaginary time evolution of $x_{L}(t)$ to be described by:
\begin{align}
	x_{L}({\rm Re}[T_{0,-}]+it_{E})\equiv x_{E}(t_{E})=-x_1\cos(\omega t_{E})-i\sqrt{x_1^2+\frac{\epsilon}{m\omega^2}}\sin(\omega t_{E})\quad;\quad 0<t_E<\frac{\pi}{\omega}
\end{align}

The Euclidean solutions, such as the above, are called instantons. The left moving solution $x_{L}(t)$, after evolution along imaginary time for a duration $i\pi/\omega$, turns into the right moving solution $x_{R}(t)$:
\begin{align}
	x_{L}(t+i\pi/\omega)=x_{R}(t)
\end{align} 
Therefore, one interprets the instanton as a bridge between the two solutions $x_{L}(t)$ and $x_{R}(t)$, which are otherwise disconnected in phase space. Hence, they offer a natural semi-classical description for quantum tunnelling and over-the-barrier reflection processes. In particular, for $\epsilon>0$, one finds that the semi-classical limit of the ratio $\mathcal{P}_{r}/\mathcal{P}_{t}$ can be written in terms of the exponential of the (Eulidean) Hamilton-Jacobi action evaluated at the instanton solution $x_{E}(t_{E})$. For the case of negative energy, a similar argument can be made for the ratio $\mathcal{P}_{t}/\mathcal{P}_{r}$. The instanton based approaches, often discussed in literature, makes use of this observation as a starting point to evaluate the semi-classical transition amplitudes. However, in the real time path integral approach, one does not have to invoke the instantons a priori. Instead, as we have seen here, it emerges naturally through complex saddle points that are relevant.

\section{Schwinger effect from real time path integral amplitudes}\label{schwinger_effect}

In this section, we shall accomplish the main objective of this paper: furnish a real time worldline path integral based representation for the positive frequency modes that give rise to Schwinger effect. Following that, we also demonstrate how the \textit{exact} value of Bogoliubov coefficients and hence, particle number $n_{\bm{k}}$ can be calculated. Let us start by looking at the classical dynamics of a charged particle interacting with an electromagnetic background. The motion is described by the following action:
\begin{align}
	\mathcal{S}_{\rm EM}=\int_{T_0}^{T}\left[\frac{m}{2}\eta_{\mu\nu}\dot{x}^{\mu}\dot{x}^{\nu}+\frac{q}{c}\dot{x}^{\mu}A_{\mu}+\frac{mc^2}{2}\right]d\tau
\end{align}
where, $\eta_{\mu\nu}=\rm{diag}(1,-1,-1,-1)$, $A_{\mu}$ is the electromagnetic gauge potential and $\tau$ is the propertime. The equation of motion that follows from varying the action is given by:
\begin{align}\label{EOM_EM}
	m\ddot{x}^{\mu}=\frac{q}{c}F^{\mu}_{\,\,\,\,\nu}\dot{x}^{\nu}
\end{align}
where, $F_{\mu\nu}$ is the electromagnetic field tensor, defined as $F_{\mu\nu}=\partial_{\mu}A_{\nu}-\partial_{\nu}A_{\mu}$. The above equation needs to be supplemented by the following constraint equation, which may be obtained by contracting the above equation with $\dot{x}_{\mu}$:
\begin{align}\label{constraint}
	\dot{x}^{\mu}\dot{x}_{\mu}=c^2
\end{align}
For a constant electric field of magnitude $E$ along, say, the x-axis, the equations of motion reduce to:
\begin{align}
	(\ddot{u},\ddot{v})&=\omega(-\dot{u},\dot{v})\quad;\quad\omega\equiv\frac{qE}{mc}\\
	m\ddot{\bm{x}}_{\perp}&=0,
\end{align}
where $u=ct-x$, $v=ct+x$ and $\bm{x}_{\perp}\equiv (y,z)$. For definiteness, we shall assume that $\omega>0$, henceforth.

If a positive frequency mode $U(x^{\mu}_1)$ can be represented as a path integral amplitude, the boundary condition at the final value of proper time, say $\tau=T$, for the trajectories that make up such a path integral, must be given by:
\begin{align}\label{xmu_final}
	x^{\mu}(T)=x_1^{\mu}=(ct_1,x_1,y_1,z_1)
\end{align}
We have to impose one more set of boundary conditions, namely, those at the initial value of proper time, say, $\tau=T_0$. To this end, we start by noting that, in the worldline picture, a positive frequency solution describes a particle moving backward in time. Following Stueckelberg \cite{ref1} and Feynman\cite{RevModPhys.20.367}, this can also be interpreted as an antiparticle moving forward in time. Therefore, the requisite initial condition should be such that it \textit{unambiguously} dispenses a solution describing a particle moving towards the negative $t$-direction. If we also assume that the particle is initially moving towards the positive $x-$direction, it is easy to show that, along the classical trajectories of this form, $\dot{u}(\tau)<0$, for all values of $\tau$ (This is analogous to $(\dot{p}-\omega \dot{x})$ being positive along classical trajectories of the IHO system). Therefore, the appropriate initial condition becomes\footnote{If, instead, we had assumed that the particle is initially moving towards the negative $x-$direction, \ref{u_initial} will be replaced by $\dot{v}(T_0)=\gamma_0<0$.}:
\begin{align}\label{u_initial}
	\left.\frac{du}{d\tau}\right|_{T_0}=\gamma_0<0
\end{align}
For convenience, we write $\gamma_0=-\omega u_0 e^{-\omega T_0}$, where $u_0>0$. Out of the remaining three initial conditions, we can use two to fix the initial momentum along the $\bm{x}_{\perp}$ directions:
\begin{align}\label{xp_initial}
	m\dot{\bm{x}}_{\perp}(T_0)=-\bm{k}_{\perp}
\end{align}  
where, $\bm{k}_{\perp}=(k_{y},k_z)$. We shall shortly fix the last remaining initial condition in a convenient way, when we look at the analysis in different gauges. The antiparticle solution $x^{\mu}_{\rm ap}(\tau)$ of  \ref{EOM_EM}, that satisfies the boundary conditions \ref{xmu_final}, \ref{u_initial} and \ref{xp_initial} is found to be:
\begin{align}
	u_{\rm ap}(\tau)&= u_1+ u_0(e^{-\omega \tau}-e^{-\omega T})\\
	v_{\rm ap}(\tau)&=v_1-\frac{1}{u_0m^2\omega^2}\left(C_0+k_{\perp}^2\right) (e^{\omega \tau}-e^{\omega T})\\
	(y_{\rm ap}(\tau),z_{\rm ap}(\tau))&=\left(y_1-\frac{k_y}{m}(\tau-T),z_1-\frac{k_z}{m}(\tau-T)\right)
\end{align} 
where, $C_0$ is an arbitrary constant, to be fixed by the remaining boundary condition and $k_{\perp}^2\equiv|\bm{k}_{\perp}|^2$. Note that we have \textit{not} imposed the constraint equation \ref{constraint} yet. Hence, in general, the solution $x^{\mu}_{\rm ap}(\tau)$ could describe time-like and space-like trajectories, when $C_0>$ and $C_0<0$, respectively. In particular, for the on-shell solutions, we obtain $C_0=m^2c^2$. One can also find a solution $x^{\mu}_{p}(\tau)$ describing a particle moving forward in time, by simply replacing $u_0\rightarrow-u_0$ in $x^{\mu}_{ap}(\tau)$, which yields:
\begin{align}
u_{\rm p}(\tau)&= u_1- u_0(e^{-\omega \tau}-e^{-\omega T})\\
v_{\rm p}(\tau)&=v_1+\frac{1}{u_0m^2\omega^2}\left(C_0+k_{\perp}^2\right) (e^{\omega \tau}-e^{\omega T})\\
(y_{\rm p}(\tau),z_{\rm p}(\tau))&=\left(y_1-\frac{k_y}{m}(\tau-T),z_1-\frac{k_z}{m}(\tau-T)\right)
\end{align} 

Motivated by our analysis of the IHO problem, we shall now modify the action $\mathcal{S}_{EM}$ by adding an appropriate boundary term that naturally imposes our initial conditions \ref{u_initial} and \ref{xp_initial}. It turns out that the relevant boundary term is dependent on the choice of gauge. In this work, we shall be concerned with three commonly used gauges, namely, the time dependent gauge: $A^{(1)}_{\mu}=(0,E ct,0,0)$, the space dependent gauge: $A^{(2)}_{\mu}=(-E x,0,0,0)$ and the lightcone gauge: $A^{(3)}_{\mu}=\frac{E}{2}(-ct-x,ct+x,0,0)$. Let us denote by $\mathcal{B}^{(i)}[x^{\mu}(T_0)]$, the appropriate boundary term corresponding to the gauge $A^{(i)}_{\mu}$, where $i=1,2,3$. Therefore, the modified action $\mathcal{S}^{(i)}_{\rm EM}$, for the gauge choice $A^{(i)}_{\mu}$, takes the form:
\begin{align}
	\mathcal{S}^{(i)}_{\rm EM}[x^{\mu};u_0,\bm{\zeta}]=\int_{T_0}^{T}\left[\frac{m}{2}\eta_{\mu\nu}\dot{x}^{\mu}\dot{x}^{\nu}+\frac{q}{c}\dot{x}^{\mu}A_{\mu}+\frac{mc^2}{2}\right]d\tau+\mathcal{B}^{(i)}[x^{\mu}(T_0);u_0,\bm{\zeta}]
\end{align} 
where, $\bm{\zeta}$ denotes the set of parameters that the modified action may depend upon. In what follows, we shall prove that the positive frequency modes $U^{(i)}(x_1^{\mu})$ in the asymptotic past, in the gauge choice $A^{(i)}_{\mu}$, can be written as the following path integral amplitude:
\begin{align}\label{PI_for_general_modes}
&U^{(i)}(x_1^{\mu})\\\nonumber
&=\int_{-\infty}^{\infty}\int^{x^{\mu}(T)=x^{\mu}_1}\exp\left\{\frac{i}{\hbar}\left[\int_{T_0}^{T}\left(\frac{m}{2}\eta_{\mu\nu}\dot{x}^{\mu}\dot{x}^{\nu}+\frac{q}{c}\dot{x}^{\mu}A_{\mu}+\frac{mc^2}{2}\right)d\tau+\mathcal{B}^{(i)}[x^{\mu}(T_0);u_0,\bm{\zeta}]\right]\right\}\mathcal{D}[x^{\mu}]\,dT
\end{align}

It is worth mentioning that a path integral formalism has already been used to study Schwinger effect in \cite{FeldbruggeJobLeon2019}. Before, moving into further details of our approach, let us review the progress made in \cite{FeldbruggeJobLeon2019} and compare it with the present work. After developing the general formalism for world line approach to Schwinger effect in a general electric field, the special case of constant electric field, which is switched on only for a finite time, is considered in \cite{FeldbruggeJobLeon2019}. Therein, using a numerical approach, they study the evolution a Gaussian wave function, with the position and momentum spread around specific values and then also visually show that the electric field facilitates a tunnelling of a particle mode to an antiparticle mode.  They also numerically investigated the different physical aspects of the wave function, that would be measured in a `weak measurement', to further establish their findings. Here, going significantly beyond the analysis in \cite{FeldbruggeJobLeon2019}, we have shown that by imposing the boundary conditions that are most natural to worldlines relevant to positive frequency modes, we can perform exact analytical computations. Such an approach has the added advantage that it makes direct contact with the standard positive frequency modes that are considered in the context of Schwinger effect. Moreover, we shall shortly apply Picard-Lefschetz theory to show how the worldline instantons emerge naturally in our formalism and explicitly see the manner in which they felicitate particle production.

\subsection{Time dependent gauge}
The boundary term appropriate to time dependent gauge turns out to be:
\begin{align}
	\mathcal{B}^{(1)}[x^{\mu}(T_0);u_0,\bm{k}]&=\frac{m\omega}{2}\left(ct(T_0)-\frac{k_x}{m\omega}-u_0e^{-\omega T_0}\right)^2+\bm{k}.\bm{x}(T_0)
\end{align} 
where, $\bm{k}=(k_x,k_y,k_z)=(k_x,\bm{k}_{\perp})$. The details of how $\mathcal{B}^{(1)}[x^{\mu}(T_0);u_0,\bm{k}]$, introduced above, naturally leads to the desired boundary conditions, viz, \ref{xmu_final}, \ref{u_initial} and \ref{xp_initial}, have been delegated to \ref{Veryfy_B_time}. We can now calculate the corresponding positive frequency mode $U_{\bm{k}}^{(1)}(x^{\mu}_1)$ using \ref{PI_for_general_modes}. After performing the exact Gaussian path integral over $x^{\mu}$ we arrive at:
\begin{align}
	U_{\bm{k}}^{(1)}(x_1^{\mu})=\int_{-\infty}^{\infty}e^{-\frac{1}{2}\omega(T-T0)}\exp\left[\frac{i}{\hbar}\mathcal{S}^{(1)}_{\rm EM}[x_{ap}^{\mu};u_0,\bm{k}]\right]dT
\end{align}
where, $x^{\mu}_{ap}$ is the classical solution that we introduced in the beginning of this section. Ignoring an overall normalization constant, we can express $U_{\bm{k}}^{(1)}(x_1^{\mu})$ in the concise form:
\begin{align}
	U_{\bm{k}}^{(1)}(x_1^{\mu})=e^{\frac{i}{\hbar}\bm{k}.\bm{x}_1}\xi^{(1)}_{\bm{k}}(t_1)
\end{align}
where,
\begin{align}\label{U1_timepart}
	\xi^{(1)}_{\bm{k}}(t_1)=\int_{-\infty}^{\infty} e^{-\frac{1}{2}\omega T}\exp\left[\frac{i}{\hbar}\tilde{\mathcal{S}}\left(T;\epsilon_{k_{\perp}},ct_1-\frac{ck_x}{qE},u_0\right)\right] dT
\end{align}
where, $\tilde{\mathcal{S}}(T;\epsilon_{k_{\perp}},x,u_0)$ is the function that we have introduced in \ref{S_epsilon}, with the energy $\epsilon$ replaced by $\epsilon_{k_\perp}=(k_{\perp}^2+m^2c^2)/(2m)$. The normalization can be fixed by demanding that the modes as orthonormal under the Klein-Gordon inner product, but, since particle production can also be studied without fixing the normalization, we shall not worry about the same henceforth. We have already seen in \ref{PI_IHO_main} that the integral in \ref{U1_timepart} can be expressed in terms of parabolic cylinder functions (see \ref{exact_psi_int} for details). Therefore, we get the final result
\begin{align}
U^{(1)}_{\bm{k}}(x^{\mu})\propto e^{\frac{i}{\hbar}\bm{k}.\bm{x}}D_{\nu_{\bm{k}}}\left[e^{\frac{3\pi i}{4}}\sqrt{\frac{2m\omega}{\hbar}}\left(ct-\frac{k_xc}{qE}\right)\right]\quad;\quad\nu_{\bm{k}}\equiv \left(-\frac{1}{2}+\frac{i\epsilon_{k_{\perp}}}{\hbar\omega}\right).
\end{align}
which is in perfect agreement with the standard result \ref{U_TD_gauge}. Similarly, one can also find the negative frequency solution in the asymptotic future, $V^{(1)*}_{-\bm{k}}(x^{\mu})$. We can spare ourselves some labour by observing that these later modes can be obtained by simply replacing $u_0\rightarrow-u_0$ in \ref{U1_timepart}.
\begin{align}\label{V1_timepart}
V_{-\bm{k}}^{(1)*}(x_1^{\mu})&=e^{\frac{i}{\hbar}\bm{k}.\bm{x}_1}\int_{-\infty}^{\infty} e^{-\frac{1}{2}\omega T}\exp\left[\frac{i}{\hbar}\tilde{\mathcal{S}}\left(T;\epsilon_{k_{\perp}},ct_1-\frac{ck_x}{qE},-u_0\right)\right] dT\\\label{define_chi}
&\equiv e^{\frac{i}{\hbar}\bm{k}.\bm{x}_1}\chi^{(1)*}_{\bm{k}}(t_1)
\end{align}
where we have, once again, ignored an overall normalization constant. The positive frequency modes of asymptotic past and asymptotic future are related by a Bogoliubov transformation of the form:
\begin{align}
	U^{(1)}_{\bm{k}}(x^{\mu})=\alpha_{\bm{k}}V_{\bm{k}}^{(1)}(x^{\mu})+\beta_{\bm{k}}V_{-\bm{k}}^{(1)*}(x^{\mu})
\end{align} 
Once the Bogoliubov coefficients $\alpha_{\bm{k}}$ and $\beta_{\bm{k}}$ are computed, the number of particles produced $n_{\bm{k}}$ can be found via $n_{\bm{k}}=|\beta_{\bm{k}}|^2$. 

\subsection{Space dependent gauge}
In the space dependent gauge, the boundary term appropriate for a particle moving backward in time and that starts towards the positive $x-$direction is given by:
\begin{align}
\mathcal{B}^{(2)}[x^{\mu}(T_0);u_0,\bm{k}_{\perp},k_t]&=-\frac{m\omega}{2}\left(x(T_0)-\frac{k_t}{cm\omega}+u_0e^{-\omega T_0}\right)^2-k_t t(T_0)+\bm{k}_{\perp}.\bm{x}_{\perp}(T_0)
\end{align}
where, $k_t>0$. The details of how $\mathcal{B}^{(2)}[x^{\mu}(T_0);u_0,\bm{k}]$, introduced above, naturally leads to the desired boundary conditions, viz, \ref{xmu_final}, \ref{u_initial} and \ref{xp_initial}, have been delegated to \ref{Veryfy_B_space}. We can now calculate the corresponding positive frequency mode $U^{(2)}_{k_t,\bm{k}_{\perp}}(x^{\mu})$ using \ref{PI_for_general_modes}. After performing the exact Gaussian path integral over $x^{\mu}$, we arrive at:
\begin{align}
U_{\bm{k}}^{(2)}(x_1^{\mu})=\int_{-\infty}^{\infty}e^{-\frac{1}{2}\omega(T-T0)}\exp\left[\frac{i}{\hbar}\mathcal{S}^{(2)}_{\rm EM}[x_{ap}^{\mu};u_0,\bm{k}_{\perp},k_t]\right]dT
\end{align}
where, $x^{\mu}_{ap}$ is the classical solution that we introduced in the beginning of this section. Ignoring an overall normalization constant, we can express $U_{k_t,\bm{k}_{\perp}}^{(2)}(x_1^{\mu})$ in the concise form:
\begin{align}
U_{k_t,\bm{k}}^{(2)}(x^{\mu})=e^{-\frac{i}{\hbar} k_t t}e^{\frac{i}{\hbar}\bm{k}_{\perp}.\bm{x}_{\perp}}\xi^{(2)}_{k_t,\bm{k}_{\perp}}(x)
\end{align}
where,
\begin{align}\label{U2_spacepart}
\xi^{(2)}_{k_t,\bm{k}_{\perp}}(x_1)=\int_{-\infty}^{\infty} e^{-\frac{1}{2}\omega T}\exp\left[-\frac{i}{\hbar}\tilde{\mathcal{S}}\left(T;-\epsilon_{k_{\perp}},x_1-\frac{k_t}{qE},-u_0\right)\right] dT
\end{align}
where, $\tilde{\mathcal{S}}(T;\epsilon_{k_{\perp}},x,u_0)$ is the function that we have introduced in \ref{S_epsilon}, with the energy $\epsilon$ replaced by $\epsilon_{k_\perp}=-(k_{\perp}^2+m^2c^2)/(2m)$. Comparing the above integral with \ref{phi_def_rightmov}, it seems that $\xi^{(2)}_{k_t,\bm{k}_{\perp}}$ is analogous to the complex conjugate of the IHO wave function $\psi_{\epsilon}^{(R)}$ that we introduced in \ref{review_iho}. There, we found that the corresponding integral evaluates to parabolic cylinder functions (the details are delegated to \ref{exact_psi_int}). Therefore, we get the final result:
\begin{align}
	U_{k_t,\bm{k}}^{(2)}(x_1^{\mu})\propto e^{-\frac{i}{\hbar} k_t t_1}e^{\frac{i}{\hbar}\bm{k}.\bm{x}_1}D_{\nu^*_{\bm{k}}}\left[e^{\frac{i\pi}{4}}\sqrt{\frac{2m\omega}{\hbar}}\left(x_1-\frac{k_t}{qE}\right)\right]
\end{align}

\subsection{The lightcone gauge}
The boundary term takes the simplest form in the lightcone gauge:
\begin{align}
	\mathcal{B}^{(3)}[x^{\mu}(T_0);u_0,\bm{k}_{\perp},k_v]=-\frac{1}{2}m\omega u_0 e^{-\omega T_0}v(T_0)-\frac{1}{2}k_\nu u(T_0)+\bm{k}_{\perp}.\bm{x}_{\perp}(T_0)
\end{align}
where, the lightcone coordinates are defined as $u=ct-x$ and $v=ct+x$. The details of how $\mathcal{B}^{(3)}[x^{\mu}(T_0);u_0,\bm{k}]$, introduced above, naturally leads to the desired boundary conditions, viz, \ref{xmu_final}, \ref{u_initial} and \ref{xp_initial}, have been delegated to \ref{Veryfy_B_light}. We can now calculate the corresponding positive frequency mode $U^{(3)}_{k_v,\bm{k}_{\perp}}(x^{\mu})$ using \ref{PI_for_general_modes}. After performing the exact Gaussian path integral over $x^{\mu}$, we arrive at:
\begin{align}
U_{\bm{k}}^{(2)}(x_1^{\mu})=\int_{-\infty}^{\infty}e^{-\frac{1}{2}\omega(T-T0)}\exp\left[\frac{i}{\hbar}\mathcal{S}^{(3)}_{\rm EM}[x_{ap}^{\mu};u_0,\bm{k}_{\perp},k_{v}]\right]dT
\end{align}
where, $x^{\mu}_{ap}$ is the classical solution that we introduced in the beginning of this section. Ignoring an overall normalization constant, we can express $U_{k_v,\bm{k}_{\perp}}^{(3)}(x_1^{\mu})$ in the concise form:
\begin{align}
U_{k_v,\bm{k}}^{(3)}(x^{\mu})=e^{-\frac{i}{2\hbar} k_v u}e^{\frac{i}{\hbar}\bm{k}_{\perp}.\bm{x}_{\perp}}\xi^{(3)}_{k_v,\bm{k}_{\perp}}(v)
\end{align}
The $v-$dependent part can further be simplified to
\begin{align}\label{U3_vpart}
\xi^{(3)}_{k_v,\bm{k}_{\perp}}(v_1)=\int_{-\infty}^{\infty} e^{-\frac{1}{2}\omega T}\exp\left[\frac{i}{\hbar}\mathcal{A}\left(T;v_1-\frac{c k_v}{qE},u_0\right)\right] dT
\end{align}
where,
\begin{align}
	\mathcal{A}\left(T;z,u_0\right)=\frac{1 }{2}m\omega z u_0e^{-\omega T}+\epsilon_{k_{\perp}}T
\end{align}
The integral in \ref{U3_vpart} can be easily recast into the standard integral representation of Gamma function, to finally arrive at:
\begin{align}
	U_{k_v,\bm{k}}^{(3)}(x^{\mu})\propto e^{-\frac{i}{2\hbar}k_v u}e^{\frac{i}{\hbar}(\bm{k}_{\perp}.\bm{x}_{\perp})}\left(1-\frac{qE}{ck_v}v\right)^{\nu_{\bm{k}}}
\end{align}
which is in perfect agreement with \ref{U3_modes_review}.
\subsection{Non-perturbative pair creation from real time path integral approach}\label{pair_production}
We have already seen in \ref{review_schwinger} that pair creation manifests itself in seemingly different ways in the three gauges that we considered here. The conventional approach to study the Schwinger effect, using canonical quantization formalism, boils down to exploiting the known properties of the standard functions in terms of which the positive/negative modes can be represented. For instance, in the time dependent gauge, one uses the connection formula of parabolic cylinder functions to read off the Bogoliubov coefficients $\alpha_{\bm{k}}$ and $\beta_{\bm{k}}$, following which, the particle number $n_{\bm{k}}$ is computed. Off course, it is also possible use the standard asymptotic methods\cite{bender2013advanced}, thereby directly arriving at the asymptotic results, without the use of special functions, which is precisely the WKB approach. The path integral approach, that we introduced here, supplies an alternate perspective on such asymptotic methods, wherein the role of `paths'(or trajectories) is significantly more transparent. Although, the manner in which particle production is inferred vary in different gauges, the algebraic aspects are essentially the same. Hence, to illustrate our point, we shall now focus only on the time dependent gauge.

In the conventional treatment of Schwinger effect, one concludes that the solutions $U^{(1)}_{\bm{k}}(x^{\mu})$ are, in fact, the positive frequency modes of the infinite past, by looking at their asymptotic limit as $t\rightarrow-\infty$. A similar analysis can also be done for $V^{(1)}_{\bm{k}}(x^{\mu})$ to verify that they realize positive frequency solutions in the infinite future. The aforementioned approach explicitly utilizes the theory of parabolic cylinder functions to accomplish the task at hand. However, the path integral representation of $U^{(1)}_{\bm{k}}(x^{\mu})$, in conjunction with Picard-Lefschetz theory, can be used to accomplish the same in a more transparent way. Let us start by looking at the asymptotic behaviour of $U^{(1)}_{\bm{k}}(x^{\mu})$ or equivalently $\xi^{(1)}_{\bm{k}}(t)$ as $t\rightarrow -\infty$. In view of \ref{psi_L_PI} and \ref{U1_timepart}, we find that $\xi^{(1)}_{\bm{k}}(t)$ has the same functional form of $\psi_{\epsilon}^{(L)}(x)$, with the identifications $\epsilon_{k_{\perp}}\leftrightarrow\epsilon$ and $(t-k_x/(qE))\leftrightarrow x$. Hence, risking a mild abuse of notations, we shall be freely lending the terminology that we used for $\psi^{(L)}_{\epsilon}(x)$ for the discussions to follow. The $t\rightarrow-\infty$ limit of $\xi^{(1)}_{\bm{k}}(t)$ follows from \ref{transmitted_Epos} and takes the form:
\begin{align}
	\xi^{(1)}_{\bm{k}}(t)\propto |t|^{\frac{i \epsilon_{k_{\perp}}}{\hbar \omega}-\frac{1}{2}}e^{\frac{ic qE t^2}{2\hbar}}\quad;\quad t\rightarrow-\infty
\end{align}
Since, in the $t\rightarrow-\infty$ limit, the phase of the above function is decreasing with respect to time $t$, we conclude that $U^{(1)}_{\bm{k}}(x^{\mu})$ describes a positive frequency solution in the asymptotic past. 

Now, in the region $(t-k_x/(qE))>0$, a convergent integral representation for $\xi^{(1)}_{\bm{k}}(t)$ can be obtained as follows:
\begin{align}\label{xi_t0_plus_t1}
	\xi^{(1)}_{\bm{k}}(t)&=\int_{\mathcal{T}_{b,+}}e^{-\frac{\omega T}{2}}\exp\left[\frac{i}{\hbar}\tilde{\mathcal{S}}(T;\epsilon_{k_{\perp}},(t-k_x/(qE)),u_0)\right]\,dT\\\nonumber
	&+ \int_{\mathcal{T}_{b,-}}e^{-\frac{\omega T}{2}}\exp\left[\frac{i}{\hbar}\tilde{\mathcal{S}}(T;\epsilon_{k_{\perp}},(t-k_x/(qE)),u_0)\right]\,dT
\end{align} 
where, $\mathcal{T}_{b,+}$ and $\mathcal{T}_{b,-}$ are, respectively, curves along steepest descent contours of the relevant saddle point $T_{0,+}$ and $T_{0,-}$ (see \ref{contours_Epos_xpos}). Next, we shall deform $\mathcal{T}_{b,-}$ to the horizontal line $\mathcal{C}_1$ shown in \ref{contours_Epos_xpos_exact}, which is defined by ${\rm Im}(\omega T)=i\pi$. Therefore, the integral along $\mathcal{T}_{b,-}$ in \ref{xi_t0_plus_t1} reduces to:
\begin{align}
	\int_{\mathcal{T}_{b,-}}e^{-\frac{\omega T}{2}}&\exp\left[\frac{i}{\hbar}\tilde{\mathcal{S}}(T;\epsilon_{k_{\perp}},(t-k_x/(qE)),u_0)\right]\,dT\\
	&=\int_{\mathcal{C}_{-}}e^{-\frac{\omega T}{2}}\exp\left[\frac{i}{\hbar}\tilde{\mathcal{S}}(T;\epsilon_{k_{\perp}},(t-k_x/(qE)),u_0)\right]\,dT\\\label{coefficient_chi}
	&=\exp\left(-i\frac{\pi}{2}-\frac{\epsilon_{k_{\perp}}}{\hbar\omega}\right)\chi^{(1)*}_{\bm{k}}(t)
\end{align}
where we have used \ref{define_chi}. Since the remaining contribution to $\xi^{(1)}_{\bm{k}}(t)$, coming from the integral along $\mathcal{T}_{b,+}$, corresponds to a wave that is purely left moving in time, it \textit{must} be proportional to $\chi^{(1)}_{\bm{k}}(t)$. This leads to:
\begin{align}
	\xi^{(1)}_{\bm{k}}(t)=\alpha_{\bm{k}}\chi^{(1)}_{\bm{k}}(t)+\beta_{\bm{k}}\chi^{(1)*}_{\bm{k}}(t)
\end{align} 
where,
\begin{align}
	\beta_{\bm{k}}=\exp\left(-i\frac{\pi}{2}-\frac{\epsilon_{k_{\perp}}}{\hbar\omega}\right)
\end{align}
from which particle number $n_{\bm{k}}=|\beta_{\bm{k}}|^2$ can be computed to get precisely \ref{particle_t_gauge}. We emphasize that the above computation of $\beta_{\bm{k}}$ (hence, of $n_{\bm{k}}$) is exact. In fact, one can also compute the \textit{exact} value of $\alpha_{\bm{k}}$ using our approach. The details of this computation can be found in \ref{alpha_derivation}. The exact value of $\alpha_{\bm{k}}$ is useful for evaluating the effective action.

\subsubsection{The emergence of instantons}
Another remarkable aspect of our formalism is the manner in which instantons emerge.  We have already seen an example of this in \ref{instanton}. In the case of Schwinger effect, as in the case of the IHO, the non-trivial quantum effects may be attributed to relevant saddle point(s) that is/are complex. In the time dependent gauge, for instance, the coefficient in front of $\xi^{(1)}_{\bm{k}}(t)$ in \ref{coefficient_chi}, which leads to particle production, arises due to the fact that $T_{0,-}$ is complex. Recall, that the final boundary condition that we introduced reads, $x^{\mu}(T)=(ct_1,x_1,y_1,z_1)$. Since the interesting features occur in the $t-x$ plane, let us consider the projection of the on-shell solution $\tilde{x}^{\mu}_{ap}(\tau)$ on to the $t-x$ plane. Then, we have:
\begin{align}
	c(\tilde{t}_{ap}(\tau)-\tilde{t}_0)&=\frac{1}{2}u_0e^{-\omega \tau}-\frac{\epsilon_{k_{\perp}}}{m\omega^2 u_0}e^{\omega \tau}\\
	(\tilde{x}_{ap}(\tau)-\tilde{x}_0)&=-\frac{1}{2}u_0e^{-\omega \tau}-\frac{\epsilon_{k_{\perp}}}{m\omega^2 u_0}e^{\omega \tau}
\end{align}   
where $\tilde{t}_0$ and $\tilde{x}_0>x_1$ are constants. Now, the complex saddle point $T_{0,-}$ can be imagined as describing the evolution of the on-shell solution along real values of proper time until $\tau={\rm Re}[T_{0,-}]$, followed by that along imaginary values of proper time for a duration ${\rm Im}[T_{0,-}]=i\pi/\omega$. Therefore, the entire evolution of an on-shell worldline corresponding to the saddle point $T_{0,-}$ takes the form:
\begin{align}\label{instanton_sol_t}
	c(\tilde{t}_{ap}(\tau)-\tilde{t}_0)&=\begin{cases}
	\frac{1}{2}u_0e^{-\omega \tau}-\frac{\epsilon_{k_{\perp}}}{m\omega^2 u_0}e^{\omega \tau}\quad&;\quad\tau<{\rm Re}[T_{0,-}]\\
-c(t_1-\tilde{t}_0)\cos(\omega\tau_{E})-i\sqrt{c^2(t_1-\tilde{t}_0)^2+\frac{\epsilon_{k_{\perp}}}{m\omega^2}}\sin(\omega\tau_{E})\quad&;\quad 0<\tau_{E}<\frac{\pi}{\omega}
	\end{cases}\\\label{instanton_sol_x}
	(\tilde{x}_{ap}(\tau)-\tilde{x}_0)&=\begin{cases}
	-\frac{1}{2}u_0e^{-\omega \tau}-\frac{\epsilon_{k_{\perp}}}{m\omega^2 u_0}e^{\omega \tau}\quad&;\quad\tau<{\rm Re}[T_{0,-}]\\
	(x_1-\tilde{x}_0)\cos(\omega\tau_{E})-i\sqrt{(x_1-\tilde{x}_0)^2-\frac{\epsilon_{k_{\perp}}}{m\omega^2}}\sin(\omega\tau_{E})\quad&;\quad 0<\tau_{E}<\frac{\pi}{\omega}
	\end{cases}
\end{align}
where, we have defined the Euclidean proper time $\tau_{E}$ through $\tau\equiv{\rm Re}[T_{0,-1}]+i\tau_{E}$ for, $0<\tau_{E}<\pi/\omega$. In fact, more generally, one could imagine the complex saddle point $T_{0,-}$ as describing a complex worldline with three parts, namely, the evolution of the on-shell antiparticle worldline (1) along real direction of proper time until $\tau={\rm Re}[T_{0,-}]-\delta T$, then (2) along imaginary direction of proper time from $\tau={\rm Re}[T_{0,-}]-\delta T$ to $\tau={\rm Re}[T_{0,-}]-\delta T+i\pi/\omega$ and finally, (3) along the real direction of proper time from $\tau={\rm Re}[T_{0,-}]-\delta T+i\pi/\omega$ to $\tau={\rm Re}[T_{0,-}]+i\pi/\omega=T_{0,-}$, where $\delta T$ is an arbitrary real number. The complex trajectories of this form are often called the wordline instantons. 

One finds, from \ref{instanton_sol_t} and \ref{instanton_sol_x}, that the (on-shell) antiparticle worldline $\tilde{x}^{\mu}_{ap}(\tau)$, after evolving along the Euclidean proper time for a duration $i\pi/\omega$, turns into the (on-shell) particle worldline $\tilde{x}^{\mu}_{p}(\tau)$. Hence, the Euclidean segment of the worldline instanton acts as a bridge, between the otherwise disconnected particle and antiparticle worldlines. Moreover, we can write the particle number $n_{\bm k}$ in terms of the action $S_{\rm EM}[\tilde{x}^{\mu}_{ap}]$ evaluated at the Euclidean segment of the worldline as:
\begin{align}\label{n_from_instanton_action}
	n_{\bm{k}}=\left|\exp\left[\frac{i}{\hbar}\int_{\textrm{Re}[T_{0,-}]}^{T_{0,-}}\left(\frac{m}{2}\eta_{\mu\nu}\dot{\tilde{x}}_{ap}^{\mu}\dot{\tilde{x}}_{ap}^{\nu}+\frac{q}{c}\dot{\tilde{x}}_{ap}^{\mu}A_{\mu}+\frac{mc^2}{2}\right)d\tau\right]\right|^2
\end{align}
Consequently, $n_{\bm k}$ can also be directly related to the probability of transition from the antiparticle worldline to a particle worldline. Most available discussions on instanton based approaches to the Schwinger effect, makes use of the observation that $n_{\bm{k}}$ can be attributed to evolution along imaginary time, right from the start. Notice, however, that we did not have to invoke the concept of instantons a priori to derive $\beta_{\bm{k}}$. On the the hand, they are found to emerge, as dictated by Picard-Lefschetz theory, from relevant saddle points that are complex. Therefore, in a way, the real time approach presents us with a more rigorous justification for the efficacy of instanton methods.   

It is worth mentioning that analytic continuations to imaginary time, usually employed in literature, are made in two distinct manners. First one corresponds to analytically continuing the Schwinger proper time $\tau$ to imaginary values, along with analytically continuing from Minkwoski to Euclidean signature for the metric. Second one corresponds to complexing only the proper time $\tau$, while retaining the Minkowski signature of the metric.  For instance, in passing from the appropriate Lorentzian worldline path integral to the Euclidean worldline path integral given \ref{Gamma_E_Euc_int}, one executes the first kind of analytic continuation\cite{Affleck:1981bma,Lavrelashvili:1989he,Dumlu:2011cc}. 
However, as is evident from our discussion so far, it is the second kind of analytic continuation that is of direct relevance to this work. For instance, one can directly see that from \ref{n_from_instanton_action} that the metric is still Minkowskian, although the proper time integral appearing in the exponential factor therein is performed along the imaginary direction. Although the two approaches are related, it is important to emphasize that they are distinct.  

\section{Conclusion} \label{conclusion}
The Schwinger effect is the simplest instance of pair creation process induced by a classical external background. Moreover, it is one of the most concrete non-perturbative predictions of quantum field theory.  Therefore, Schwinger effect has been a subject of intense theoretical and experimental investigation for decades. Several approaches, of differing mathematical and physical implications, have been proposed for the theoretical study of this problem. However, the more conventional approaches, based on the standard tools of QFT, seldom offer a satisfactory reconciliation with our point particle intuitions. On the other hand, the available literature on worldline path integral approach, which offers elegant and more intuitive interpretations for the pair creation process, often invoke imaginary time a priori. In this work, we proposed a Lorentzian worldline path integral approach to Schwinger effect, which is, by construction, based on sum over worldlines of a charged particle, evolving in \textit{real} proper time. 

In particular, we have explicitly shown that the standard Schwinger modes, in different gauges, can expressed as worldline path integral amplitudes. It is well known that a positive frequency solution, in the worldline formalism, translates to a particle moving backward in time. This condition, in turn, can be mathematically imposed on the worldine by an initial condition given by \ref{u_initial}, which is also supplemented by the initial conditions that correspond to fixing momenta in different spatial directions. In order to naturally implement the aforementioned initial conditions in the classical variational problem, we modified the standard worldline action by introducing appropriate boundary terms. Remarkably, this simple procedure leads to the positive frequency Schwinger modes for all the three gauges that we considered, despite the fact that mode functions have different mathematical forms in each choice of gauge. 

Following this, using our formalism, we derived the \text{exact} values of Bogoliubov coeffiects and, hence, the particle number. We emphasize that, for the purpose of this calculations, we did not have to explicitly use \textit{any} knowledge of the theory of parabolic cylinder functions. To rephrase it, the path integral approach affords a means to quantitatively compute/infer pair creation, even if we had, hypothetically, no knowledge of the representation of positive/negative Schwinger modes in terms of standard functions. This is remarkable, especially when we consider the fact that, in a general electromagnetic field configuration, we do not have the luxury of having the analytical form of mode functions at our disposal. In light of this, we believe that exploration of pair creation in more complicated backgrounds using real time path integral approach is a promising avenue. It would also be interesting to see how one can extent our approach to study the backreaction of produced pairs, on the electromagnetic field. Another avenue worth exploring would be to seek how subtler aspects of the problem, like renormalization and the effect of internal photons, may be dealt with in the Lorentzian path integral approach. These topics will be explored in a future publication.

\section*{Acknowledgement}
K.R. gratefully acknowledges several helpful discussions with Thanu Padmanabhan and Sumanta Chakraborty. K.R. is supported by the Research Associateship of Indian Association for the Cultivation of Science (IACS), Kolkata, India.

\appendix	
\labelformat{section}{Appendix #1}
\labelformat{subsection}{Appendix #1}
\labelformat{subsubsection}{Appendix #1}
\section{Exact computation of the IHO scattering wave function}\label{exact_psi_int}
We start by shifting the real-time integration contour in \ref{phi_def_2}, to the horizontal line $\textrm{Im}[\omega T]=-i\pi/4$, so that we get a more convergent integral:
\begin{align}\label{scatter_ww_def_2}
\psi^{(L)}_{\epsilon}(x_1)&=\mathcal{N} \int_{-\infty-i\frac{\pi}{4}}^{\infty-i\frac{\pi}{4}}e^{-\frac{\omega T}{2}}\exp\left[\frac{i}{\hbar}\tilde{\mathcal{S}}(T;\epsilon,x_1,u_0)\right]\,dT
\end{align}  
The above integral can be transformed into one of the standard integral representations of the parabolic cylinder functions by the following change of variable:
\begin{align}
s\equiv \sqrt{\frac{m\omega}{2\hbar}}u_0e^{-\left(\omega T+i\frac{\pi}{4}\right)}
\end{align}  
so that the integral in \ref{scatter_ww_def_2} takes the form:
\begin{align}
\psi^{(L)}_{\epsilon}(x_1)&\propto \,\,e^{-\frac{z^2}{4}}\int_{0}^{\infty}e^{-zs-\frac{s^2}{2}}s^{-(\nu+1)}\quad;\quad z=e^{\frac{3\pi i}{4}}\sqrt{\frac{2m\omega}{\hbar}}x\\\label{psi_L_para}
&\propto D_{\nu}\left(e^{\frac{3\pi i}{4}}\sqrt{\frac{2m\omega}{\hbar}}x_1\right)
\end{align}
which is in perfect agreement with \ref{psi_L_first_appear}. Similarly, one can also obtain the `right-moving' solution $\psi^{(R)}_{\epsilon}(x_1)$ by replacing $u_0\rightarrow-u_0$ in the right hand side of \ref{phi_def_2}, which yields:  
\begin{align}\label{phi_def_rightmov2}
\psi^{(R)}_{\epsilon}(x_1)&\propto \int_{-\infty}^{\infty}e^{-\frac{\omega T}{2}}\exp\left[\frac{i m \omega}{2}\left(x_1+u_0 e^{-\omega T}\right)^2-\frac{i m u_0^2 \omega }{4}e^{-2\omega T }+i\epsilon T\right]\,dT\\
&\propto D_{\nu}\left(e^{-\frac{\pi i}{4}}\sqrt{\frac{2m\omega}{\hbar}}x_1\right)
\end{align}
where, $u_0>0$ and the proportionality constant in the last line of the above equation is the same as that in \ref{psi_L_para}.

\section{Variation of the modified actions}\label{action_verify}
In this section we shall verify that the action $\mathcal{S}^{(i)}_{\rm EM}[x^{\mu};u_0,\bm{\zeta}]$ naturally furnishes a variational problem for the boundary conditions \ref{xmu_final}, \ref{u_initial} and \ref{xp_initial}, when the choice of gauge is $A^{(i)}_{\mu}(x^{\mu})$.
\subsection{Time dependent gauge}\label{Veryfy_B_time}
Consider the variation of action $\mathcal{S}^{(1)}_{\rm EM}$.
\begin{align}
\delta \mathcal{S}^{(1)}_{\rm EM}&=\int_{T_0}^{T}\left(-m\ddot{x}_{\mu}+F_{\mu\nu}\dot{x}^{\nu}\right)\delta x^{\mu} d\tau+mc\dot{t}(T)\delta t(T)-\left[m\dot{x}(T)-m\omega c t(T)\right]\delta x(T)\\\nonumber
&+\left[-mc\dot{t}(T_0)+m\omega ct(T_0)-k_x-m\omega u_0e^{-\omega T_0}\right]\delta t(T_0)+\left[m\dot{x}(T_0)-m\omega ct(T_0)+k_x\right]\delta x(T_0)\\\nonumber
&-m\dot{\bm{x}}_{\perp}(T).\delta\bm{x}_{\perp}(T)+\left[m\dot{\bm{x}}_{\perp}(T_0)+\bm{k}_{\perp}\right].\delta\bm{x}_{\perp}(T_0)
\end{align}
Hence, when $x^{\mu}(T_0)$ are not fixed, the action $\mathcal{S}^{(1)}_{\rm EM}$ furnishes a well defined variational principle for the following boundary conditions:
\begin{align}\label{T_gauge_BC_1}
x^{\mu}(T)&=(ct_1,x_1,y_1,z_1)\\\label{T_gauge_BC_2}
-mc\dot{t}(T_0)+m\omega ct(T_0)-k_x-m\omega u_0e^{-\omega T_0}&=0\\\label{T_gauge_BC_3}
m\dot{x}(T_0)-m\omega ct(T_0)+k_x&=0\\\label{T_gauge_BC_4}
m\dot{\bm{x}}_{\perp}(T_0)+\bm{k}_{\perp}&=0
\end{align} 
It is easy to verify the above boundary conditions retain \ref{xmu_final}, \ref{u_initial} and \ref{xp_initial}. The remaining boundary condition is supplied by \ref{T_gauge_BC_3}, which effectively fixes the arbitrary constant $C_0$, in the antiparticle solution $x^{\mu}_{ap}(\tau)$, to:
\begin{align}
C_0=m u_0\omega  e^{-2\omega T } \left[2e^{\omega T } (k_x-m \omega  ct_1)+m u_0 \omega \right]-k_{\perp}^2
\end{align}
Therefore, the constraint equation, which imposes $C_0=m^2c^2$, essentially fixes the value of $T$ to its on-shell value. 
\subsection{Space dependent gauge}\label{Veryfy_B_space}
The variation of action $\mathcal{S}^{(2)}_{\rm EM}$ takes the form:
\begin{align}
\delta \mathcal{S}^{(2)}_{\rm EM}&=\int_{T_0}^{T}\left(-m\ddot{x}_{\mu}+F_{\mu\nu}\dot{x}^{\nu}\right)\delta x^{\mu} d\tau+[mc\dot{t}(T)-m\omega x(T)]\delta t(T)-m\dot{x}(T)\delta x(T)\\\nonumber
&+\left[-mc\dot{t}(T_0)+m\omega x(T_0)-\frac{k_t}{c}\right]\delta t(T_0)+\left[m\dot{x}(T_0)-m\omega x(T_0)+\frac{k_t}{c}-m\omega u_0e^{-\omega T_0}\right]\delta x(T_0)\\\nonumber
&-m\dot{\bm{x}}_{\perp}(T).\delta\bm{x}_{\perp}(T)+\left[m\dot{\bm{x}}_{\perp}(T_0)\bm{k}_{\perp}\right].\delta\bm{x}_{\perp}(T_0)
\end{align}
Hence, with $x^{\mu}(T_0)$ kept free to vary, the action $\mathcal{S}_{\rm EM}^{(2)}$ furnishes a well defined variational problem for the following boundary conditions:
\begin{align}\label{X_gauge_BC_1}
x^{\mu}(T)&=(ct_1,x_1,y_1,z_1)\\\label{X_gauge_BC_2}
-mc\dot{t}(T_0)+m\omega x(T_0)-\frac{k_t}{c}&=0\\\label{X_gauge_BC_3}
m\dot{x}(T_0)-m\omega x(T_0)+\frac{k_t}{c}-m\omega u_0e^{-\omega T_0}&=0\\\label{X_gauge_BC_4}
m\dot{\bm{x}}_{\perp}(T_0)+\bm{k}_{\perp}&=0
\end{align} 
A straightforward algebra shows that the above boundary conditions retain \ref{xmu_final}, \ref{u_initial} and \ref{xp_initial}. The remaining boundary condition, on the other hand, is supplied by \ref{X_gauge_BC_2}, the effect of which is to fix the arbitrary constant $C_0$, in the antiparticle solution $x^{\mu}_{ap}(\tau)$, to:
\begin{align}
C_0=m u_0\omega  e^{-2 T \omega } \left(2 e^{T \omega } \left(\frac{k_t}{c}-m x_1 \omega \right)-m u_0 \omega \right)-k_{\perp}^2
\end{align}
Therefore, the constraint equation, which imposes $C_0=m^2c^2$, essentially fixes the value of $T$ to its on-shell value. 
\subsection{Lightcone gauge}\label{Veryfy_B_light}
To verify that $\mathcal{B}^{(3)}[x^{\mu}(T_0);u_0,\bm{k}_{\perp},k_v]$ is the desired boundary condition in lightcone gauge, let us look at the variation of the corresponding action:
\begin{align}
	\delta \mathcal{S}^{(3)}_{\rm EM}&=\int_{T_0}^{T}\left(-m\ddot{x}_{\mu}+F_{\mu\nu}\dot{x}^{\nu}\right)\delta x^{\mu} d\tau+[mc\dot{t}(T)-m\omega x(T)]\delta t(T)-m\dot{x}(T)\delta x(T)\\\nonumber
	&+\frac{1}{2}\left[-m\dot{u}(T_0)-m\omega u_0 e^{-\omega T_0}\right]\delta v(T_0)+\frac{1}{2}\left[-m\dot{v}(T_0)+m\omega v(T_0)-k_v\right]\delta u(T_0)\\\nonumber
	&-m\dot{\bm{x}}_{\perp}(T).\delta\bm{x}_{\perp}(T)+\left[m\dot{\bm{x}}_{\perp}(T_0)\bm{k}_{\perp}\right].\delta\bm{x}_{\perp}(T_0)
\end{align}
\begin{align}\label{LC_gauge_BC_1}
x^{\mu}(T)&=(ct_1,x_1,y_1,z_1)\\\label{LC_gauge_BC_2}
\dot{u}(T_0)+\omega u_0 e^{-\omega T_0}&=0\\\label{LC_gauge_BC_3}
-m\dot{v}(T_0)+m\omega v(T_0)-k_v&=0\\\label{LC_gauge_BC_4}
m\dot{\bm{x}}_{\perp}(T_0)+\bm{k}_{\perp}&=0
\end{align}
Hence, $\mathcal{S}^{(3)}$ effects the desired boundary conditions, namely, \ref{xmu_final}, \ref{u_initial} and \ref{xp_initial}, in a natural way, with the remaining initial condition given by \ref{LC_gauge_BC_3}. The effect of the latter is to fix the arbitrary constant $C_0$ in the classical solution $x_{ap}(x^{\mu})$ to the value:
\begin{align}
C_0=m u_0 \omega  e^{-T \omega } (k_v-m v_1 \omega )-k_{\perp}^2
\end{align}
Once again, we find that the constraint equation, which imposes $C_0=m^2c^2$, essentially fixes the value of $T$ to its on-shell value. 
\section{Derivation of $\alpha_{k}$}\label{alpha_derivation}
Here, we present a method to derive the \textit{exact} value of $\alpha_{\bm{k}}$, without directly using the theory of parabolic cylinder functions. Since the following argument holds true for any contour $\mathcal{C}_+$ that can be continuously deformed into $\mathcal{T}_{b,+}$, we have also plotted such a curve for representational purpose in \ref{contours_Epos_xpos_exact}. As we have already seen in \ref{pair_production}, the (time dependent part of ) positive frequency mode $\xi^{(1)}_{\bm{k}}(t)$ can be written as:
\begin{align}\label{xi_t0_plus_t1_v2}
\xi^{(1)}_{\bm{k}}(t)&=\int_{\mathcal{T}_{b,+}}e^{-\frac{\omega T}{2}}\exp\left[\frac{i}{\hbar}\tilde{\mathcal{S}}(T;\epsilon_{k_{\perp}},(t-k_x/(qE)),u_0)\right]\,dT+\beta_{\bm{k}}\chi^{(1)*}_{\bm{k}}(t)
\end{align}    
The first term on the right hand side \textit{must} be proportional to $\chi^{(1)}_{\bm{k}}(t)$, since the same has contribution from the saddle point $T_{0,+}$ only. Hence, by definition, the Bogoliubov coefficient $\alpha_{\bm{k}}$ is given by
\begin{align}
	\alpha_{\bm{k}}=\frac{\int_{\mathcal{T}_{b,+}}e^{-\frac{\omega T}{2}}\exp\left[\frac{i}{\hbar}\tilde{\mathcal{S}}(T;\epsilon_{k_{\perp}},(t-k_x/(qE)),u_0)\right]\,dT}{\chi^{(1)}_{\bm{k}}(t)}
\end{align}
Since the value of $t$ at which the right hand side of the above equation is evaluated does not matter, we might as well fix it to a value of our convenience, say, $t=\frac{k_x}{qE}$. The expression for $\alpha_{\bm{k}}$ reduces to:
\begin{align}
	\alpha_{\bm{k}}=\frac{\int_{\mathcal{T}_{b,+}}e^{-\frac{\omega T}{2}}\exp\left[\frac{i}{\hbar}\tilde{\mathcal{S}}(T;\epsilon_{k_{\perp}},0,u_0)\right]\,dT}{\int_{-\infty}^{\infty} e^{-\frac{1}{2}\omega T}\exp\left[-\frac{i}{\hbar}\tilde{\mathcal{S}}\left(T;\epsilon_{k_{\perp}},0,-u_0\right)\right] dT}
\end{align}
To evaluate the numerator, we introduce the variable $s=i\sqrt{(m\omega u^2_0)/(4\hbar)}e^{-2\omega T}$, so that
\begin{align}
	\int_{\mathcal{T}_{b,+}}e^{-\frac{\omega T}{2}}\exp\left[\frac{i}{\hbar}\tilde{\mathcal{S}}(T;\epsilon_{k_{\perp}},0,u_0)\right]\,dT&=\left[\frac{1}{2\omega}\left(\frac{4\hbar}{m\omega u_0^2i}\right)^{-\frac{\nu_{\bm{k}}}{2}}\right]\int_{-\infty-i0^{+}}^{-\infty+i0^+}e^{s}s^{-\frac{\nu_{\bm{k}}}{2}-1}ds\\
	&=\left[\frac{1}{2\omega}\left(\frac{4\hbar}{m\omega u_0^2i}\right)^{-\frac{\nu_{\bm{k}}}{2}}\right]\frac{2\pi i}{\Gamma\left(\frac{\nu_{\bm{k}}}{2}+1\right)}
\end{align}
where, we have used the following standard integral representations of the Gamma function:
\begin{align}
	\frac{2\pi i}{\Gamma(\beta)}=\int_{-\infty-i0^{+}}^{-\infty+i0^+}e^{s}s^{-\beta}ds
\end{align}
A similar change of variable can be used to simplify the denominator to obtain:
\begin{align}
	\int_{-\infty}^{\infty} e^{-\frac{1}{2}\omega T}\exp\left[-\frac{i}{\hbar}\tilde{\mathcal{S}}\left(T;\epsilon_{k_{\perp}},0,-u_0\right)\right] dT=\left[\frac{1}{2\omega}\left(\frac{4\hbar}{m\omega u_0^2 i}\right)^{\frac{\nu_{\bm{k}}+1}{2}}\right]\Gamma\left(\frac{\nu_{\bm{k}}+1}{2}\right)
\end{align}
Therefore, the final expression for $\alpha_{\bm{k}}$ becomes:
\begin{align}
	\alpha_{\bm{k}}=\frac{-\sqrt{2\pi}e^{i\frac{\pi \nu_{\bm{k}}}{2}}\exp\left[i\left(\frac{\gamma_{\bm{k}}\epsilon_{k_{\perp}}}{\hbar\omega}\right)\right]}{\Gamma(1+\nu_{\bm{k}})}
\end{align}
where, $\gamma_{\bm{k}}=\log(2\hbar/(mu_0^2\omega))$, which, interestingly, vanishes for $u_0=2\hbar c/(qE)$.  It is easy to verify that the Bogoliubov coefficients that we obtained satisfy $|\alpha_{\bm{k}}|^2-|\beta_{\bm{k}}|^2=1$. 
\bibliography{Path_integral_Schwinger_effect}

\bibliographystyle{utphys1}
\end{document}